\documentclass[11pt,a4paper,notitlepage]{article}
\usepackage[utf8]{inputenc}
\usepackage[english]{babel}
\usepackage[T1]{fontenc}
\usepackage{hyperref}
\usepackage{amsmath}
\usepackage{amssymb}
\usepackage{amsfonts}
\usepackage{slashed}
\usepackage{color} 
\usepackage{caption}
\usepackage{subcaption}
\usepackage{soul}
\usepackage{cite}
\usepackage{array,multirow,makecell}
\setcellgapes{1pt}
\makegapedcells
\newcolumntype{R}[1]{>{\raggedleft\arraybackslash }b{#1}}
\newcolumntype{L}[1]{>{\raggedright\arraybackslash }b{#1}}
\newcolumntype{C}[1]{>{\centering\arraybackslash }b{#1}}
\newcommand{\Tr}{\mathrm{Tr}}
\newcommand{\tr}{\mathrm{tr}}

\newcommand{\inv}{\mathrm{inv}}

\usepackage{enumerate}
\usepackage{tikz-cd}

\newcommand{\cB}{{\mathcal B}}

\newcommand{\cF}{{\mathcal F}}
\newcommand{\cG}{{\mathcal G}}

\newcommand{\cJ}{{\mathcal J}}

\newcommand{\cT}{{\mathcal T}}

\definecolor{mygray}{gray}{0.3}

\newcommand\beq{\begin{equation}}
\newcommand\eeq{\end{equation}}
\newcommand{\bes}{\begin{eqnarray}}
\newcommand{\ees}{\end{eqnarray}}
\def\nn{{\nonumber}}
\newcommand{\one}{\mbox{$1 \hspace{-1.0mm}  {\bf l}$}}
\def\inv{{\mbox{\tiny -1}}}

\def\vphi{{\varphi}}

\newcommand\restr[2]{{
  \left.\kern-\nulldelimiterspace 
  #1 
  \vphantom{\big|} 
  \right|_{#2} 
  }}

\def\extd{\mathrm {d}}
\newcommand{\U}{\mathrm{U}}
\newcommand{\SU}{\mathrm{SU}}
\usepackage{multicol}
\usepackage{amssymb}
\usepackage{graphicx}
\usepackage[left=2cm,right=2cm,top=2.5cm,bottom=2.5cm]{geometry}

\begin{document}
\begin{center}
\textbf{\Large{Tensor models and group field theories: combinatorics, large $N$ and renormalization
}}

\vspace{15pt}

{\large Sylvain Carrozza$^{a,}$\footnote{\url{sylvain.carrozza@u-bourgogne.fr}}}

\vspace{10pt}

$^{a}${\sl Institut de Math\'{e}matiques de Bourgogne, UMR 5584, CNRS \& Universit\'{e} de Bourgogne, F-21000 Dijon, France.
}


\end{center}

\vspace{5pt}

\begin{abstract}
\noindent We provide a brief overview of tensor models and group field theories, focusing on their main common features. Both frameworks arose in the context of quantum gravity research
, and can be understood as higher-dimensional generalizations of matrix models. We describe the common combinatorial structure underlying such models and review some of the key mathematical results that have been obtained in this area. Of central importance is the discovery of a new family of large $N$ expansions for random tensors. It has driven applications of tensor models to random geometry and non-perturbative local quantum field theory, and has also enabled the development of rigorous renormalization methods for group field theories and other non-local quantum field theories. 
\end{abstract}


\section{Introduction}

With the pioneering work of t' Hooft \cite{tHooft:1973alw}, it was recognized that a quantum field theory (QFT) of an $N \times N$ matrix field can be formally expanded into Feynman amplitudes indexed by \emph{ribbon graphs} or \emph{combinatorial maps} \cite{lando2004graphs}. Those two equivalent combinatorial structures formalize the notion of a graph embedded into a canonically associated surface (considered up to homeomorphism), and 
the genus of that surface indexes a \emph{topological large $N$ expansion} in which $\frac{1}{N^2}$ takes the role of a formal small parameter. In addition to t' Hooft's original idea of using this expansion to probe the non-perturbative sector of QFT, it has
found numerous applications to enumerative combinatorics, and more specifically, to two-dimensional quantum gravity \cite{Brezin:1977sv, David:1984tx, Ambjorn:1985az, DiFrancesco:1993cyw}. Indeed, formal matrix integrals 
can be used to efficiently manipulate generating functions of combinatorial maps, and asymptotic enumeration results for such objects can be inferred by studying the critical properties of the underlying matrix model \cite{Eynard:2015aea, Eynard:2016yaa}. From a physical point of view, such a critical regime can be interpreted as defining the continuum limit of a two-dimensional random geometry \cite{le2013uniqueness, miermont2013brownian} which is equivalent to Liouville quantum gravity \cite{miller2015liouville, gwynne2020random}. 

Tensor models were first considered in the early nineties \cite{Ambjorn:1990ge, Sasakura:1990fs, Gross:1991hx}, in an attempt to establish an analogous correspondence between random tensors\footnote{By convention, we reserve the word \emph{tensor} for an object with $D \geq 3$ indices, to contrast with the case of vectors ($D=1$) and matrices ($D=2$).} and Euclidean quantum gravity 
in dimension $D \geq 3$. These early works pointed to a fertile correspondence between the Feynman expansion of such a tensor model (indexed by higher generalizations of ribbon graphs) and discrete geometries in dimension $D$. However, no generalization of the large $N$ expansion of matrix models could be identified at the time. For a while, this hindered further progress on tensor models and their relationship to random geometry.  

In the meantime was introduced the formalism of Group Field Theory (GFT) \cite{Oriti:2006se, Freidel:2005qe, Krajewski:2011zzu}. A GFT can be understood as an algebraically enriched 
tensor model which  generates, via its Feynman expansion, discrete spaces weighted by lattice gauge theory amplitudes (rather than mere combinatorial factors). More precisely, it is a (perturbative) QFT of a tensor field whose indices take value into a group $G$
, and the action is chosen in such a way as to generate the amplitudes of a specific lattice gauge theory with structure group $G$. Particularly studied types of actions are those that generating $BF$ lattice gauge theory amplitudes \cite{Boulatov:1992vp, Ooguri:1992eb} and discrete quantum gravity amplitudes in holonomy-flux variables \cite{DePietri:1999bx, Reisenberger:2000zc, Baratin:2011hp, Baratin:2011tx}. 
The latter can be interpreted as giving rise to models of random geometry, in which the geometry is function of both the combinatorial structure of the lattice and its attached group-theoretic data. From a mathematical point of view, GFTs suffer from divergences which are of a similar nature as those encountered in local perturbative QFT, but owing to the non-local character of a GFT action, are beyond the scope of standard perturbative QFT theorems. Hence, for some time, it was not clear if and in which sense GFTs could be considered as mathematically well-defined perturbative QFTs.

In 2009, the formalism of \emph{colored} tensor models (resp. GFTs) was introduced by Gurau \cite{Gurau:2009tw}, which eventually led to the discovery of a new large $N$ expansion \cite{Gurau:2010ba, Gurau:2011aq, Gurau:2011xq}. The availability of this crucial ingredient, which generalizes the large $N$ expansion of matrix models but also differs from it in crucial respects, triggered new developments in both tensor models and GFTs. On the one hand, the large $N$ expansion has allowed to make precise statements about the relation of random tensors to random geometry (see Sec.~\ref{sec:random_geo}). On the other hand, the combinatorial tools developed in the context of tensor models have enabled the proof of a series of renormalization results adapted to GFT (and other non-local QFTs), which have considerably clarified their mathematical definition (see Sec.~\ref{sec:GFT_ren}). Finally, Gurau's large $N$ expansion and its variants (see Sec.~\ref{sec:largeN}) were later exploited to explore the non-perturbative sector of local QFTs with internal symmetry described by tensor representations (see Sec.~\ref{sec:local_tft}). See \cite{Gurau:2024nzv} for a recent review.

Our review is organized as follows. In Sec.~\ref{sec:TM}, we describe Gurau's large $N$ expansion of tensor models\footnote{Gurau's first paper on the large $N$ expansion was aimed at both tensor models and GFTs, but further results in that direction were for the most part focused on the former. 
} and some of its variants, together with their two main applications to date: random geometry and strongly-coupled QFT. GFTs and closely related families of non-local QFTs are the focus of Sec.~\ref{sec:GFT}. 
Regarding GFT, we restrict our attention to renormalization results since they have a strong connection to tensor models. Other aspects of GFT, which have received continued attention and led to important developments in recent years 
are only briefly mentioned in Sec.~\ref{subsec:GFT-cosmo}. 


\section{Tensor models}\label{sec:TM}

\subsection{Combinatorial structure and melonic large \textit{N} expansions}\label{sec:largeN}


In its simplest incarnation, a tensor model takes as input a linear subspace $\cT$ of multilinear maps acting on some (real, complex or quaternionic) vector space of dimension $N$, together with the natural group action of the appropriate unitary group (namely, ${\rm U}(N)$, ${\rm O}(N)$ or ${\rm Sp}(N)$).\footnote{We will see shortly that it is advantageous to consider more constraining symmetries defined by a \emph{product} of such unitary groups.} If $\cT$ is a space of tensors of order $p$, one chooses an appropriate basis and simply represents $T \in \cT$ by a multidimensional array $T_{a_1 \cdots a_p}$ ($a_1\,, \ldots a_p \in \{1 , \ldots , N\}$). One can then equip $\cT$ with a formal measure taking the form:
\begin{equation}\label{eq:measure}
    \extd \nu (T) = \extd \mu_{\pmb P} (T) \, \exp\left( - S_{\rm int}(T) \right) \,, \qquad S_{\rm int}(T) = \sum_k \lambda_k I_k (T)\,,
\end{equation}
where: $\pmb P$ is a symmetric linear operator on $\cT$, and $\extd \mu_{\pmb P}$ the Gaussian measure of covariance $\pmb P$; the (interaction part of the) action $S_{\rm int}(T)$ is a polynomial invariant of the tensor $T$, which is typically expanded in terms of a basis of invariant monomials $\{ I_k (T) \}$; the coupling constants $\{ \lambda_k \}$ are understood as formal perturbative parameters. In all the models considered in the present section, $\pmb P$ is itself assumed to be equivariant under the group action on $\cT$ (see Sec.~\ref{sec:GFT} for models in which this condition is relaxed). One is then interested in determining the universal properties of such an object in the asymptotic limit of large dimensionality, namely $N \to + \infty$. With a few notable exceptions (e.g. \cite{Gurau:2011kk, Gurau:2013pca}), this question has been investigated in the sense of formal power-series in $\{ \lambda_k \}$. For this reason, we will make no attempt to provide a probabilistic definition of the formal measure $\extd \nu (T)$, and will allow ourselves to consider any kind of invariant in the action $S_{\rm int}(T)$ (e.g. real invariants which are unbounded from below). What makes tensor models interesting from this formal point of view is that quantities such as the free energy $F \propto \ln \int \extd \nu (T)$ can be interpreted as generating functions of combinatorial objects, namely combinatorial maps and higher-dimensional generalizations thereof.

\begin{figure}
    \centering
    \includegraphics{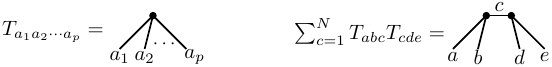}
    \caption{Graphical representation of tensors and their contractions.}
    \label{fig:tensors_contractions}
\end{figure}

\begin{figure}
    \centering
    \includegraphics[scale=.8]{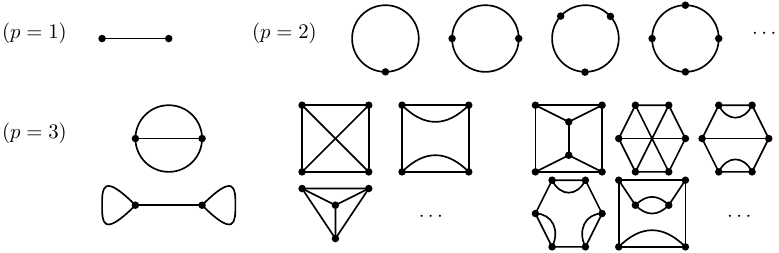}
    \caption{Connected invariants (organized according to their polynomial order) in the case of: a real vector ($p=1$); a Hermitian matrix ($p=2$); and a real symmetric tensor of order $3$ $(p=3)$.}
    \label{fig:invariants}
\end{figure}

An important observation is that the space of polynomial tensor invariants is significantly richer for $p \geq 3$ than it is when $p\leq 2$. To see this, it is convenient to represent tensors and their index-contractions graphically, as illustrated in Fig.~\ref{fig:tensors_contractions}: a tensor of order $p$ is represented by a vertex connected to $p$ half-edges, while the contraction of two indices amounts to gluing the two corresponding half-edges into an edge. Basic polynomial invariants can then be represented by closed graphs, which we illustrate in Fig.~\ref{fig:invariants} for a real vector $\phi$ ($p=1$), an Hermitian matrix $M$ ($p=2$), and a completely symmetric real tensor of order $T$ ($p=3$). In all three cases, a generic polynomial invariant can be decomposed as a sum of products of more basic invariants, which can be represented by connected graphs, and are thus called \emph{connected invariants}. But the spaces of connected invariants are very different in each case. The vector $\phi$ admits a single connected invariant: the inner product $\phi_a \phi_a$, represented by a single edge.\footnote{From now on, Einstein's convention of implicit summation over repeated indices is assumed.} By contrast, the matrix $M$ admits an infinite set of connected invariants, one per order $n \in \mathbb{N}$: the $n^{\rm th}$ power sum of its eigenvalues, $\tr(M^n)$, which is represented by a cyclic graph with $n$ vertices. Furthermore, any two such invariants are independent in the limit $N \to +\infty$. Finally, the connected invariants of a real symmetric tensor $T$ of order $3$ are in one-to-one correspondence with $3$-regular (multi)graphs, whose number grows super-exponentially with the number $n$ of vertices (or, equivalently, the order of the invariant). 
Again, any two such connected invariants become independent when $N$ is sufficiently large. Such a super-exponential growth of the number of connected invariants 
is observed for any $p \geq 3$ and for other types of tensors.\footnote{See e.g.~\cite{BenGeloun:2013lim, BenGeloun:2020lfe} for precise enumeration results.} It makes the theory space of tensor models comparatively more difficult to analyze than those of vector or matrix models. However, as it turns out, the generating functions produced by specific families of tensor invariants can be of a simpler nature than those of matrix models, while remaining richer than those of vector models, which is in part what makes tensor models physically interesting. 

\medskip

\begin{figure}
    \centering
\begin{subfigure}{.4\textwidth}
    \centering
    \includegraphics[scale =.8]{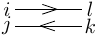}
    \caption{Propagator.}
\end{subfigure}%
\begin{subfigure}{.4\textwidth}
    \centering
    \includegraphics[scale =.8]{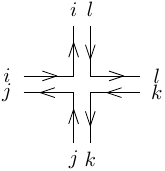}
    \caption{Vertex.}
\end{subfigure}
    \caption{Propagator and interaction vertex of the quartic Hermitian matrix model: an oriented strand between matrix indices $i$ and $j$ represents a Kronecker $\delta_{ij}$.}
    \label{fig:propa_vertex_matrix}
\end{figure}

A standard example of matrix model can be obtained by taking $\cT$ to be the vector space of $N\times N$ Hermitian matrices, while setting $\pmb P = \frac{1}{N} \one$ (where $\one$ is the identity map on $\cT$) and $S(T) =-\frac{\lambda}{4} N \tr(T^4)$ in equation \eqref{eq:measure}. The primary objects of interest in such a model are expectation values of invariant observables relative to the measure $\extd \nu$, which can in principle be reconstructed from correlation functions of the form $\langle \tr(T^{n_1})\cdots \tr(T^{n_q}) \rangle$. The perturbative expansion in $\lambda$ of such correlators can in turn be organized in terms of Feynman graphs, which in this example are $4$-regular ribbon diagrams: they can be obtained by consistently gluing an arbitrary number of propagators and vertices with the structure shown in Fig.~\ref{fig:propa_vertex_matrix}. Crucially, such ribbon diagrams can be canonically interpreted as quadrangulations of surfaces with boundaries, and the $N$-dependence of their Feynman amplitude is entirely determined by the topology of the surface. Focusing on the free energy, which generates closed and connected quadrangulations, we obtain the celebrated topological genus expansion:
\begin{equation}\label{eq:topological_exp}
\ln \int_\cT \extd\nu(T) = \sum_{g \in \mathbb{N}} N^{2-2g} \cF_g (\lambda)\,,
\end{equation}
where $\cF_g (\lambda)$ is the generating function of connected quadrangulations of genus $g$ (and $\lambda$ counts the number of quadrangles). In the asymptotic limit $N \to \infty$, such a matrix model is therefore characterized by the planar sector $g=0$, and finite $N$ corrections can in principle be investigated in a perturbative expansion in $1/N^2$. Interestingly, the structure of such an expansion  allows the genus-$g$ free energy $\cF_g (\lambda)$ to admit better convergence properties than the full free energy, which enables a non-perturbative treatment of the coupling constant $\lambda$ at fixed genus (and in particular in the planar sector). This was the original motivation of 't Hooft for devising such a large $N$ expansion in the context of large $N$ gauge theory. Finally, the topological expansion \eqref{eq:topological_exp} can be straighforwardly generalized to actions of the form $S(T) = N \tr(V(T))$, where $V$ is some ($N$-independent) polynomial. Including a monomial proportional to $T^n$ in $V$ amounts to allowing ribbon vertices of degree $n$, or dually, faces of degree $n$ in the discrete surfaces generated by the model.

\medskip

\begin{figure}
    \centering
\begin{subfigure}{.4\textwidth}
    \centering
    \includegraphics[scale=.7]{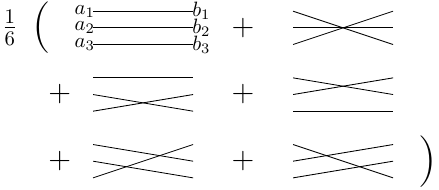}
    \caption{Propagator ${\pmb P}_{a_1 a_2 a_3 , b_1 b_2 b_3}$.}
\end{subfigure}%
\begin{subfigure}{.4\textwidth}
    \centering
    \includegraphics[scale=.9]{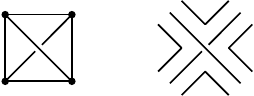}
    \caption{'Tetrahedral' invariant and its kernel.}
\end{subfigure}
    \caption{Propagator and interaction vertex of the quartic symmetric tensor model.}
    \label{fig:sym_tens}
\end{figure}

A natural generalization of the previous model to the tensor realm can be obtain by letting $\cT$ be the space of real tensors of size $N^3$ (equipped with its standard inner product), and $\pmb P$ the orthogonal projector onto its symmetric subspace.\footnote{This is equivalent to taking $\cT$ to be the space of completely symmetric real tensors and $\pmb P$ the identity, but is somewhat more convenient for our purpose.} A possible generalization of the quartic interaction we considered in the matrix case is given by the invariant represented in Fig.~\ref{fig:sym_tens}, which is often times called 'tetrahedral' due to its combinatorial stucture (we will come back to this). This leads to the action:
\begin{equation}\label{eq:action_sym}
    S_{\rm int}(T)= \frac{-\lambda}{4N^\alpha}  \, T_{abc} T_{dec} T_{dbf} T_{aef}\,, 
\end{equation}
where $\lambda$ is the perturbative parameter and $\alpha$ is a scaling parameter that will need to be set to an appropriate value in order to ensure the existence of a large $N$ expansion.\footnote{Note that the tetrahedral invariant is real and unbounded in both directions, contrary to the positive quartic interaction we were using in the matrix context.} The model we have just defined was proposed in the very first papers on tensor models \cite{Ambjorn:1990ge, Sasakura:1990fs, Gross:1991hx} as a way to generate, via the Feynman expansion, three-dimensional 'discrete spaces' obtained by gluing an arbitrary numbers of tetrahedra along their faces, in analogy with matrix models. What is for sure is that the Feynman graphs of the model can be interpreted as higher generalizations of ribbon diagrams, in which through each propagator run three strands instead of two; see Fig.~\ref{fig:sym_tens}. It is however more challenging to provide a precise definition of what one would mean exactly by 'discrete space' here
. The interplay between the combinatorial structure of the action and the global structure of the Feynman diagrams, is indeed much less rigid in higher dimensions than it is in two dimensions: it is therefore challenging to ensure nice global properties (such as a manifold structure) from the Feynman rules (which only define local gluing conditions). Even more problematic, no analogue of the topological large $N$ expansion of equation \eqref{eq:topological_exp} could be identified in this model. Progress on these questions was first achieved in the context of so-called \emph{colored} tensor models, in which more constraining symmetries make the combinatorial structure of the Feynman diagrams more rigid. We now turn to a description of this important class of models, before briefly returning to real symmetric tensors at the end of the section.     

\medskip 

An interesting structural feature of colored tensor models is their relation to the combinatorial species of \emph{edge-colored graphs}. This relation was first exploited in the context of models involving a multiplet of complex tensors \cite{Gurau:2009tw, Gurau:2010nd, Bonzom:2011zz, Gurau:2011xp}, whose entries were labeled by an index called \emph{color}. These were later subsumed by so-called \emph{uncolored} models \cite{Bonzom:2012hw}, which are somewhat simpler to describe in that they involve a single complex random tensor, and have as a result imposed themselves as a better starting point to develop a general theory of random tensors \cite{gurau2017random}. We will focus on this latter class of models here, which we will simply refer to as \emph{colored} tensor models.\footnote{Due to later developments, the historical distinction between \emph{colored} and \emph{uncolored} models can be confusing.}

\begin{figure}
    \centering
    \includegraphics[scale=.8]{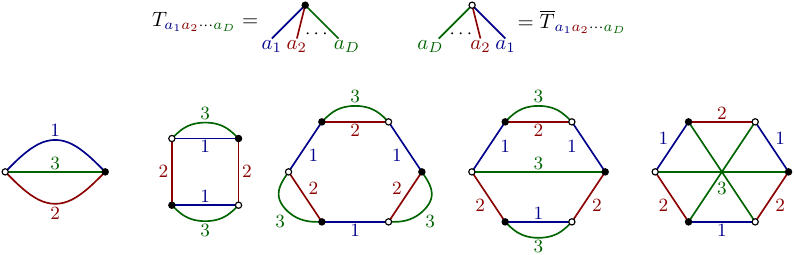}
    \caption{Representation of colored tensor invariants as $D$-colored graphs $(D=3)$.}
    \label{fig:Un_tensors}
\end{figure}

\begin{figure}
    \centering
\begin{subfigure}{.4\textwidth}
    \centering
    \includegraphics{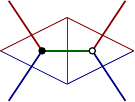}
    \caption{$D=3$.}
\end{subfigure}%
\begin{subfigure}{.4\textwidth}
    \centering
    \includegraphics{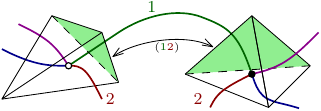}
    \caption{$D=4$.}
\end{subfigure}
    \caption{Interpretation of $D$-colored graphs as dual $(D-1)$-pseudomanifolds.}
    \label{fig:colored_dual}
\end{figure}

\begin{figure}
    \centering
    \includegraphics[scale=.7]{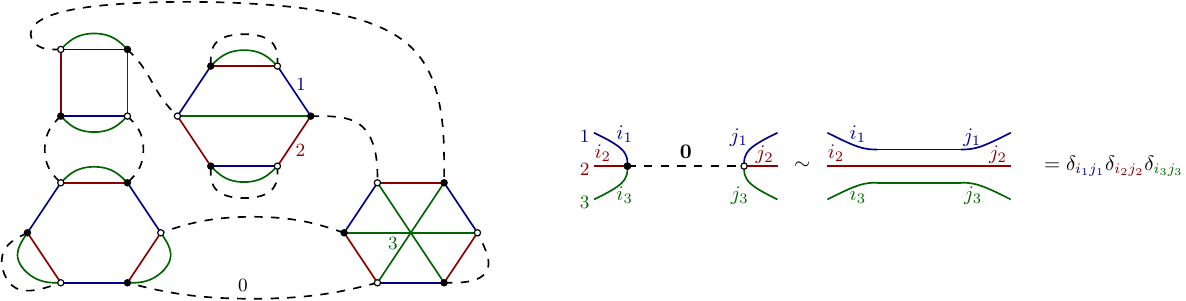}
    \caption{A Feynman diagram represented as a $(D+1)$-colored graph ($D=3$), where color-$0$ edges are dashed to emphasize they represent propagators (by contrast, other colored edges are internal to interaction vertices). It can be reinterpreted as a stranded graph upon substituting color-$0$ edges by stranded propagators, as shown in the right panel, and its amplitude is determined by interpreting each strand as a Kronecker delta.}
    \label{fig:Feynman_colored}
\end{figure}

A \emph{complex colored tensor model} \cite{gurau2017random} is a theory of a complex random tensor $T_{a_1 \cdots a_D}$ of size $N^D$, living in the $D$-fundamental representation of $\mathrm{U}(N)^D$. This means there is one independent unitary symmetry for each index of $T$: for any $U^{(1)}, \ldots , U^{(D)} \in \mathrm{U}(N)$, we have the transformation law
\begin{equation}\label{eq:sym_Un}
    T_{a_1 \cdots a_D} \to U^{(1)}_{a_1 b_1} \cdots U^{(D)}_{a_D b_D} T_{b_1 \cdots b_D}\,.
\end{equation}
The formal measure is defined as in equation \eqref{eq:measure} with the covariance taken to be the identity, so that
\begin{equation}\label{eq:colored_sym}
    \int \extd \mu_{\pmb P} (\bar T , T) \bar T_{a_1 \cdots a_D} T_{a_1 \cdots b_D} = \prod_{k = 1}^D \delta_{a_k b_k}\,,
\end{equation}
while the action is assumed to be an invariant of the form:
\begin{equation}\label{eq:action_colored}
    S_{\rm int}(\bar T , T) = - \sum_\cB \lambda_\cB N^{- \alpha(\cB)} \Tr_\cB (\bar T , T)\,.
\end{equation}
In the last formula, the sum is taken over (a finite set of) \emph{connected and bipartite $D$-colored graphs} $\{ \cB \}$, which index a generating set of invariants $\{ \Tr_\cB (\bar T , T)  \}$ under the group action \eqref{eq:sym_Un}; see Fig.~\ref{fig:Un_tensors} for some examples. Connected $D$-colored graphs, also known as \emph{bubbles} in this context, have two types of vertices of coordination $D$ (black and white), representing a tensor $T$ or its conjugate $\bar T$, and $D$ types of edges labeled by an index in $\{ 1, \ldots, D\}$, called \emph{color}. An edge labeled by $k \in \{ 1, \ldots, D\}$ always connects a black to a white vertex, and represents the invariant contraction of two indices in position $k$. By a construction originally due to Pezzana which matured into a combinatorial approach to PL manifolds known as \emph{crystallization theory} \cite{ferri1986graph} (see also \cite{Gurau:2011xp, lionni2018colored}), any bubble diagram admits a dual interpretation as a colored triangulation of an orientable \emph{pseudomanifold}\footnote{This is a generalization of the notion of topological manifold that allows mild types of topological singularities.} of dimension $D-1$.\footnote{This topological interpretation is the reason why the number of indices of a colored random tensor is denoted by the letter $D$.} For instance, the first four bubbles shown in Fig.~\ref{fig:Un_tensors} represent triangulations of the $2$-sphere, while the fifth is dual to a triangulation of the $2$-torus. As soon as $D\geq 3$, the colored structure of the bubbles is essential to their dual topological interpretation \cite{Gurau:2010nd}: the color labels allow to unambiguously specify how subsimplices of arbitrary co-dimension should be identified, which is illustrated in the right panel of Fig.~\ref{fig:colored_dual}. Taking the topological cone over each such $(D-1)$-dimensional pseudomanifold yields an elementary building block of $D$-dimensional space (a $3$-ball in the first four examples of Fig.~\ref{fig:Un_tensors}, a pseudomanifold with a pointlike singularity in the fifth example). This construction provides a precise higher-dimensional generalization of the correspondence between matrix invariants and $n$-gons (which is sharper, and therefore more satisfactory than what we had in the symmetric tensor model with action \eqref{eq:action_sym}). In the Feynman expansion, colored invariants are glued together by propagator edges, which we can assign a new color label ($0$ by convention). With this convention in place, Feynman diagrams have the structure of $(D+1)$-colored graphs.  They are therefore dual to pseudomanifolds of dimension $D$, which are precisely obtained by gluing together the previously mentioned topological cones along their boundaries. Having the same type of combinatorial structure encoding both tensor invariants and Feynman graphs is a remarkable property of colored tensor models. See Fig.~\ref{fig:Feynman_colored} for an example of Feynman graph, and for its reinterpretation as a (colored) stranded diagram. Finally, coming back to equation \eqref{eq:action_colored}, the parameters $\{ \alpha(\cB) \}$ need to be set to appropriate values in order to ensure the existence of a large $N$ limit. This choice is not unique, but we will focus here on the so-called \emph{melonic} large $N$ scaling \cite{Bonzom:2012hw}, which is governed by a combinatorial quantity known as the \emph{Gurau degree} $\omega$ \cite{Gurau:2010ba, Gurau:2011aq, Gurau:2011xq, Gurau:2011xp}. 

Given a connected $(D+1)$-colored graph $\cG$ (e.g. a Feynman diagram), let $p(\cG) \in \mathbb{N}$ be its number of white (resp. black) vertices and $F(\cG)$ its number of \emph{faces}, defined as the number of bi-colored cycles in $\cG$. The Gurau degree of $\cG$ is a \emph{positive integer}, defined by \cite{Gurau:2010ba, Gurau:2011aq}
\begin{equation}\label{eq:degree}
   \omega(\cG) = D - F(\cG) + \frac{D(D-1)}{2} p(\cG) \,. 
\end{equation}
The positivity of $\omega$ is not explicit in this formula, but can be recognized by re-expressing \eqref{eq:degree} as \cite{Gurau:2011xq}
\begin{equation}\label{eq:jackets}
   \omega(\cG) = \frac{1}{(D-1)!} \sum_{\sigma} g(\cJ(\cG,\sigma))\,,
\end{equation}
where $\sigma$ runs over cyclic permutations of $\{ 0 , \ldots , D\}$, $\cJ(\cG,\sigma)$ is a combinatorial map canonically associated to the pair $(\cG , \sigma)$ -- dubbed \emph{jacket} of $\cG$ in \cite{BenGeloun:2010wbk} --, and $g(\cJ(\cG,\sigma)) \in \mathbb{N}$ is its topological genus. The positivity of $\omega$ makes it an interesting higher-dimensional analogue of the standard genus of an orientable surface. This is reinforced by the realization that $\omega$ is nothing but the orientable genus when $D=2$ (which can be seen from both \eqref{eq:degree} and \eqref{eq:jackets}). However, an important difference between $D=2$ and $D\geq 3$ is that $\omega$ is \emph{not a topological invariant} when $D \geq 3$. See \cite{Gurau:2011xp, gurau2017random, lionni2018colored} for further detail. 

\begin{figure}
    \centering
\begin{subfigure}{.4\textwidth}
    \centering
    \includegraphics[scale=.7]{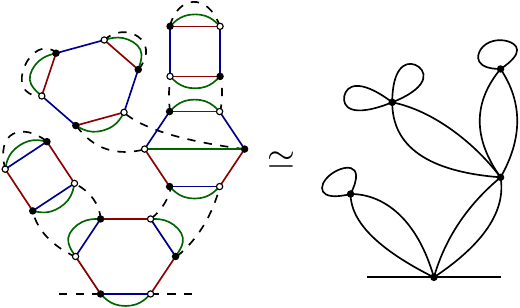}
    \caption{}
\end{subfigure}%
\begin{subfigure}{.4\textwidth}
    \centering
    \includegraphics[scale=.7]{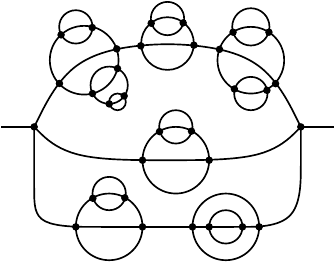}
    \caption{}
\end{subfigure}
    \caption{Two families of melonic diagrams: (a) 'local' melons from complex colored models; (b) 'bilocal' melons from models with 'tetrahedral' interaction, as e.g. in the real $\mathrm{O}(N)^3$ model (each vertex in the picture represents a 'tetrahedral' interaction, whose internal structure has been omitted).}
    \label{fig:local_bilocal}
\end{figure}

Now, if one sets\footnote{A bubble $\cB$ being a connected $D$-colored graph, its degree is defined by \eqref{eq:degree} after substituting $D$ for $D-1$.} 
\begin{equation}\label{eq:melonic_scaling}
\alpha(\cB) := \frac{2}{(D-2)!} \omega(\cB)   
\end{equation}
for any bubble $\cB$ appearing in equation \eqref{eq:action_colored}, one can prove the existence of a sensible large $N$ expansion generalizing the one of matrix models \cite{Bonzom:2012hw}. In particular, one finds the following expression for the partition function
\begin{equation}\label{eq:largeN_complex_colored}
    \ln \int \extd \nu(\bar T , T ) = \sum_{\omega \in \mathbb{N}} N^{D - \frac{2}{(D-1)!}\omega} \cF_\omega (\{ \lambda_\cB \}) \,,
\end{equation}
where $\cF_\omega (\{ \lambda_\cB \})$ is the generating function of connected $(D+1)$-colored graph, enumerated by number of bubbles of each type (the variable $\lambda_\cB$ counting the number of bubbles of type $\cB$). The leading order sector consists in connected $(D+1)$-colored graphs of vanishing degree, which are known as \emph{melonic graphs} \cite{Bonzom:2011zz}. This is the analogue of the planar sector of matrix models, and any melonic graph is indeed mapped via Pezzana's theorem to a triangulation of the $D$-sphere. However, since $\omega$ is not topological in $D\geq3$, not every colored triangulation of the $D$-sphere is melonic. In fact, the set of melonic graphs is quite restricted: it can be generated recursively, from  the unique colored graph on two vertices, by a finite set of elementary graph insertions along edges.\footnote{The "melonic" qualifier is motivated by the shape of such an elementary insertion, which is reminiscent of a cataloupe melon; see Fig.~\ref{fig:local_bilocal}.} Melonic graphs are therefore in bijection with combinatorial trees (more specifically $(D+1)$-ary trees). 
This will become important when we discuss applications of melonic large $N$ expansions. 

\medskip


Other large $N$ limits dominated by families of melonic diagrams (as we will see, with slightly different combinatorial properties) have been obtained in closely related tensor models. These include the original colored tensor model, built from a $(D+1)$-tuple of complex tensors \cite{Gurau:2010ba, Gurau:2011aq, Gurau:2011xq}. In the context of models involving a single tensor, there is some freedom in the choice of symmetry group, while preserving the colored structure of the Feynman graphs. 
Early on, the so-called \emph{multi-orientable} tensor model \cite{Tanasa:2011ur, Dartois:2013he, Raasakka:2013eda, Fusy:2014rba, Tanasa:2015uhr}, involving a complex $3$-tensor, was shown to admit a melonic limit with an action proportional to a 'tetrahedral' interaction of the form $\bar T_{abc} T_{dec} \bar T_{dbf} T_{aef}$ (note the similarity with \eqref{eq:action_sym}). The $\U(N) \times \mathrm{O}(N) \times \U(N)$ invariance of this model paved the way to a more systematic exploration of \emph{real colored tensor models} obeying $\mathrm{O}(N)^D$ invariance. The $D=3$ case was first considered in \cite{Carrozza:2015adg} (with the first two interactions shown in Fig.~\ref{fig:On_quartic_invariants}), and later generalized to higher $D$ in \cite{Ferrari:2017jgw}. At the combinatorial level, the main difference between real and complex colored models is that real invariants (resp. real Feynman diagrams) are labeled by connected $D$-colored graphs (resp. $(D+1)$-colored graphs) with \emph{no bipartite restriction} imposed
; see Fig.~\ref{fig:On_quartic_invariants} for some examples. In particular, for $D=3$, the action \eqref{eq:action_sym} obeys the desired $\mathrm{O}(N)^3$ symmetry and can rightly be represented as a non-bipartite $3$-colored graph ($\cB_T$ in Fig.~\ref{fig:On_quartic_invariants}). The latter can be interpretated under Pezzana's duality as a triangulation of a non-orientable pseudomanifold.\footnote{More precisely, it has the topology of the real projective plane. From this point of view, it is therefore somewhat misleading to call it 'tetrahedral', which explains our use of quotation marks.} Focusing on the $\mathrm{O}(N)^3$ model with interaction \eqref{eq:action_sym} for simplicity, one can prove that, with the large $N$ scaling defined by setting $\alpha = \frac{3}{2}$, a large $N$ expansion exists. In particular, the partition function expands as \cite{Carrozza:2015adg, Bonzom:2019moj}
\begin{equation}\label{eq:largeN_On}
    \ln \int \extd \nu(T ) = \sum_{\omega \in \frac{\mathbb{N}}{2}} N^{3 - \omega} \cF_\omega (\lambda) \,,
\end{equation}
where $\omega$ is a combinatorial quantity associated to Feynman graphs -- again called \emph{degree} --, and $\cF_\omega (\lambda)$ is the generating function of connected $4$-colored graphs with degree $\omega$. The degree of a graph $\cG$ is defined as $\omega(\cG):=3 + 3p(\cG) - F(\cG)$ (which agrees with \eqref{eq:degree}), and can again be re-expressed as a sum of genera of ribbon graphs canonically associated to $\cG$. However, owing to the absence of bipartite structure in the Feynman graphs, those ribbon graphs do not need to be orientable, which explains why $\omega$ is in general a positive \emph{half}-integer. As in complex models, leading order Feynman graphs ($\omega = 0$) have a melonic structure, albeit of a slightly different type: from the point of view of the Feynman expansion, a melonic insertion has a \emph{bilocal} structure, while melonic decorations where always \emph{local} (that is, tadpole-like) in complex models. This difference is illustrated in Fig.~\ref{fig:local_bilocal}.  Generalizing this construction to arbitrary $D > 3$ is not that straightforward, but interactions supporting a 'bilocal' melonic large $N$ expansion have been identified for any \emph{prime} $D \geq 3$ in \cite{Ferrari:2017jgw, Valette:2019nzp}: they are associated to certain \emph{complete} colored graphs (the 'tetrahedral' interaction being itself represented by a complete graph on four vertices), and the somewhat surprising prime condition on $D$ is the result of combinatorial constraints that need to be imposed on the colorings of such complete graphs
. 

Finally, colored tensor models with more than one asymptotic parameter have been discussed in the literature. For instance, reference \cite{Ferrari:2017ryl} introduced a $3$-index random tensor obeying $\U(N) \times \mathrm{O}(D) \times \U(N)$,\footnote{$D$ here has nothing to do with the number of indices of the tensor.} which can be analyzed in a double asymptotic expansion in the (formally) small parameters $1/N^2$ and $1/\sqrt{D}$. This  model can equivalently be understood as a colored multi-matrix model involving a large number ($D \to +\infty$) of matrices. The so-called \emph{large $D$ expansion} of this model \cite{Ferrari:2017ryl, Azeyanagi:2017mre} was extended to higher order tensors in \cite{Ferrari:2017jgw}, and a double-scaling limit in which $\frac{N^2}{D}$ is kept fixed was studied in \cite{Benedetti:2020iyz, Bonzom:2022yvc, Avohou:2023wvg}. An interesting aspect of these models is that they allow to embed a tensor-like melonic expansion of the form \eqref{eq:largeN_On} into the more standard genus expansion of a matrix model.

\medskip


Finally, we briefly come back to the model described around equation \eqref{eq:action_sym}, whose symmetry was described by a single copy of $\mathrm{O}(N)$ acting on a completely symmetric real tensor. This early proposal received renewed attention after it was convincingly argued in \cite{Klebanov:2017nlk} that it should support a (bilocal) melonic large $N$ expansion if the random tensor is not only taken to be symmetric, but also \emph{traceless}. This crucial observation, supported by explicit empirical checks, was proven rigorously in \cite{Benedetti:2017qxl} (see also \cite{Gurau:2017qya}), then generalized to the other two \emph{irreducible} $3$-tensor representations of $\mathrm{O}(N)$ \cite{Benedetti:2017qxl, Carrozza:2018ewt}. These results were then extended to $5$-index irreducible $\mathrm{O}(N)$ tensors ($p=5$) \cite{Carrozza:2021qos}, which suggests that bilocal melonic limits might exist for arbitrary $p \geq 3$ (or, at the very least, prime $p$). 
Similar techniques have allowed to partially extend the results of \cite{Ferrari:2017ryl} about the large $D$ expansion of colored matrix models to Hermitian matrices \cite{Carrozza:2020eaz}. In all those works (and particularly so in \cite{Benedetti:2017qxl} and \cite{Carrozza:2021qos}), the absence of colored structure makes the combinatorial proofs traditionally employed in the analysis of tensor models rather involved. This calls for the development of alternative methods, which might be better suited to the analysis of irreducible random tensors. Finally, tensor models with symplectic symmetry have been considered in \cite{Carrozza:2018psc} ($\mathrm{Sp}(N)$) and \cite{Gurau:2022dbx, Keppler:2023lkb} ($\mathrm{Sp}(2N, \mathbb{R})$).
 
\subsection{Random geometry}\label{sec:random_geo}

It was realized in the 70s and 80s that matrix models can be used to model continuum random surfaces, or 2d Euclidean quantum gravity \cite{Brezin:1977sv, David:1984tx, Ambjorn:1985az, DiFrancesco:1993cyw}. In a first step, the topological large $N$ expansion 
is used to generate 
an ensemble of random discretizations of two-dimensional orientable surfaces as in \eqref{eq:topological_exp}. Taking the strict $N\to +\infty$ limit of this partition function (appropriately rescaled by $1/N^2$) produces a generating function $\cF_0 (\lambda)$ of discretizations of the $2$-sphere. Each such discretization can be endowed with a natural metric,\footnote{For instance, one can attribute the same flat reference metric to all building blocks, and in this way endow the surface with a piecewise flat metric. In the dual ribbon graph representation, one may instead rely on the graph distance.} so one can think of the matrix model as producing a random discrete metric space. A (somewhat heuristic) continuum limit can then be obtained by tuning $\lambda$ to the (real, positive) critical value $\lambda_c$ of $\cF_0$.\footnote{This makes sense because $\cF_0$ indeed has a finite radius of convergence, even though the full partition function one started from had vanishing radius of convergence.} The basic idea is that, as $\lambda \to \lambda_c^-$ the sum over surfaces becomes dominated by discrete geometries involving an arbitrarily large number of elementary building blocks
. The upshot is that the \emph{critical regime}  of matrix integrals in the $N\to +\infty$ limit may be thought of as encoding continuum random surfaces of genus $0$.\footnote{As a side note, a non-trivial sum over topology can be recovered by performing a so-called \emph{double-scaling limit}, in which the $N \to +\infty$ and $\lambda \to \lambda_c$ limits are performed jointly while keeping the quantity $N(\lambda - \lambda_c)^{\alpha}$ fixed, for some appropriate $\alpha > 0$.} The probabilistic object behind this limit, known as the \emph{Brownian map} \cite{le2013uniqueness, miermont2013brownian}
, has been constructed rigorously as the $n \to + \infty$ limit of various families of uniform random planar maps with fixed number $n$ of vertices.\footnote{Each map is equipped with the graph distance, suitably rescaled by $n^{-1/4}$ in the $n\to +\infty$ limit.} The result is a random metric space which is (almost surely) homeomorphic to a $2$-sphere, has spectral dimension $2$ and Hausdorff dimension $4$.\footnote{Convergence is defined in the sense of Gromov-Hausdorff in the space of all compact metric spaces.} Furthermore, the limit is universal in the sense that it is largely independent of the details of the combinatorial maps being considered (which in the matrix model picture, is related to the choice of potential), and it is furthermore equivalent to Liouville quantum gravity \cite{miller2015liouville, gwynne2020random}. 


\medskip 

Random tensor models were initially motivated by the challenge of generalizing the success story of matrix models to Euclidean quantum gravity in $D\geq 3$ dimension. Of all attempts made until now, the framework of complex colored tensor models seems to be the soundest from a topological point of view, thanks to its connection to colored diagrams, which can be rigorously interpreted as higher-dimensional discretized (pseudo)manifolds.\footnote{In $D=2$, it was proven that uniform colored triangulations of the $2$-sphere do converge to the Brownian map \cite{carrance2021convergence}, so we have no a priori reason to reject colored triangulations as an insufficiently general combinatorial structure to explore random geometry in $D\geq 3$.} Moreover, the large $N$ limit of \eqref{eq:largeN_complex_colored} generates an ensemble of colored triangulations of the $D$-sphere, as was the case in $D=2$. The main difference is that these are limited to melonic triangulations, which constitute a quite restricted set of triangulated spheres. The fact that they are in bijection with trees actually leads to a rather simple behaviour as compared to what is found in matrix models. For instance, the generating function $G(\lambda)$ of rooted melonic $(D+1)$-colored triangulations, which governs the genus-$0$ sector of the original colored tensor model (as well as an 'uncolored' version \eqref{eq:action_colored} in which an infinite set of bubble interactions have been switched on \cite{Bonzom:2012hw}), obeys the simple polynomial equation:
\begin{equation}\label{eq:fuss-catalan}
    G(\lambda) = 1 + \lambda G(\lambda)^{D+1}\,,
\end{equation}
which one may recognize as the equation for the generating function of Fuss-Catalan numbers. The critical properties of $G$, as well as the asymptotic enumeration of (rooted) melonic $(D+1)$-colored triangulations, can readily be extracted from this equation \cite{Bonzom:2011zz}, leading to
\begin{equation}
    G(\lambda) \underset{\lambda \to \lambda_c^-}{\sim} K \sqrt{ \lambda_c - \lambda } \,, \qquad \# \{ \mathrm{rooted}\;\mathrm{colored}\;\mathrm{triangulations}\;\mathrm{of}\;\mathrm{size}\;n\} \underset{n \to +\infty}{\sim} K \lambda_c^n n^{-3/2}\,.
\end{equation}
The square-root singularity of $G(\lambda)$ and the exponent $-3/2$ in the second relation are directly related \cite{flajolet2009analytic},\footnote{By contrast, the number of rooted planar maps with $n$ vertices scales asymptotically with an exponent $-5/2$, which is typical of planar combinatorial objects.} and indicative of a continuum limit governed by Aldous' \emph{continuous random tree} \cite{10.1214/aop/1176990534, Aldous_1991} (also known as the \emph{branched polymer phase} in the physics literature \cite{Ambjorn:1990wp, Bialas:1996ya}). This is a random metric space of non-integer spectral dimension $4/3$ (and Hausdorff dimension $2$), which can be obtained as a scaling limit of uniform random trees of size $n$ (equipped with the graph distance multiplied by a corrective factor $n^{-1/2}$). It was confirmed in \cite{Gurau:2013cbh} that uniform melonic triangulations of size $n$ similarly converge to Aldous' continuum random tree, \emph{for any topological dimension} $D \geq 3$. This is a somewhat disappointing but rather remarkable universality result: while tensor models in the melonic scaling \eqref{eq:melonic_scaling} offer an interesting generalization of the topological large $N$ expansion of matrix models, unlike the latter, the random geometries they generate tend to degenerate into tree-like metric spaces in the continuum limit. More generally, it is reasonable to expect that the same result should hold for any other model exhibiting a large $N$ limit dominated by tree-like combinatorial species. Henceforth, a number of attempts have been made to evade such tree-like combinatorial behaviour, which we now briefly describe. 



\medskip


First, double-scaling limits in which the large $N$ and critical limits are taken simultaneously have been constructed \cite{Kaminski:2013maa, Dartois:2013sra, Bonzom:2014oua}, with the idea that incorporating subleading contributions of tensor models into the continuum limit might be sufficient to change its universality class. However, the structure of fixed-degree Feynman  graph, as unraveled in the complex colored setting in \cite{gurau2016regular}, seems to invalidate this strategy for the standard melonic scaling of \eqref{eq:melonic_scaling}: for any $\omega \in \mathbb{N}$, there is only a finite number of connected Feynman graphs of degree $\omega$ modulo insertions of melonic two-point subgraphs and ladder four-point subgraphs (such equivalent classes of graphs are known as \emph{schemes}). Given that the generating function of ladder diagrams can be resummed explicitly as a geometric series (see Fig.~\ref{fig:sde}), there seems to be no room to evade the universality class of branched polymers. Furthermore, similar results have been obtained within alternative melonic large $N$ expansions, such as e.g. the $\mathrm{O}(N)^3$ model \cite{Fusy:2014rba, Gurau:2015tua, Bonzom:2019yik, Bonzom:2021kjy, Krajewski:2023tbv}. 


However, as argued in Sec.~\ref{sec:largeN}, the theory space of tensor models is much vaster than that of matrix models. In particular, other scalings than the melonic one can be considered to obtain alternative but consistent large $N$ limits.\footnote{In fact, due to the proliferation of invariants with arbtrarily complex combinatorial structures, determining the optimal scaling of a bubble $\cB$ (i.e. the smallest value of $\alpha(\cB)$ in \eqref{eq:largeN_complex_colored} compatible with the existence of a large $N$ expansion) is a challenging question in general. See \cite{lionni2018colored} and references therein for further detail.} For instance, it is straightforward to emulate matrix-like large $N$ limits in the context of a tensor model with $D$ even \cite{Bonzom:2015axa}, which demonstrates that the Brownian map universality class can be obtained within tensor models. More complicated examples of non-melonic large $N$ limits have been constructed in \cite{Bonzom:2015axa, bonzom2017counting, Lionni:2017xvn, lionni2018colored}, but all of them lead to a critical regime which is either tree-like or planar-like, or that mixes tree-like and planar-like structures. In $D=3$, a theorem due to Bonzom \cite{Bonzom:2018btd} indicates that any reasonable bubble interaction (that is, a $3$-colored graph of genus $g=0$, representing the triangulated boundary of a $3$-ball) will lead to the universality class of branched polymers. 


Further strategies might be considered. For instance, if one attempts to generate random geometries from other tensor model observables than the generating function of connected Feynman graphs, the universality class of the continuum limit might change. Indeed, it was observed in the colored multi-matrix model of \cite{Benedetti:2020iyz} (considered in some triple-scaling limit) that restricting the sum over graphs to $2$-particle irreducible (or, equivalently, $3$-edge connected) contributions was sufficient to turn a melonic critical regime into a model of random maps coupled to the Ising model. Whether such a  combinatorial mechanism might prove useful to generate a higher-dimensional random geometry from random tensors remains to be explored. 


For now, getting out of the universality classes of continuous random trees and random maps remains a challenge for tensor models. More broadly, obtaining a non-trivial  random metric space with integer spectral dimension $D\geq 3$ as a scaling limit of discrete geometries is an outstanding open question for any approach; see e.g. the extensive literature on (causal) dynamical triangulations (\cite{Ambjorn:2024pyv} and references therein), or \cite{Lionni:2019bzb, Budd:2022dhq} for more specific proposals aiming at overcoming this roadblock. 


\subsection{Large \textit{N} (local) quantum field theory}\label{sec:local_tft}

What limits the scope of melonic large $N$ limits in the purely combinatorial context of random geometry (a universal tree-like behaviour at criticality) makes them quite useful in the context of large $N$ quantum mechanics and QFT. More precisely, the \emph{bilocal} melonic limits described in Sec.~\ref{sec:largeN} have been taken advantage of to explore the strong coupling regime of Q(F)T by means of analytical methods \cite{Delporte:2018iyf, Klebanov:2018fzb, Gurau:2019qag, Benedetti:2020seh}, a feat which is typically out of reach even at large $N$ (at least in the absence of additional structures, such as integrable ones).

\begin{figure}
    \centering
    \includegraphics[scale=.8]{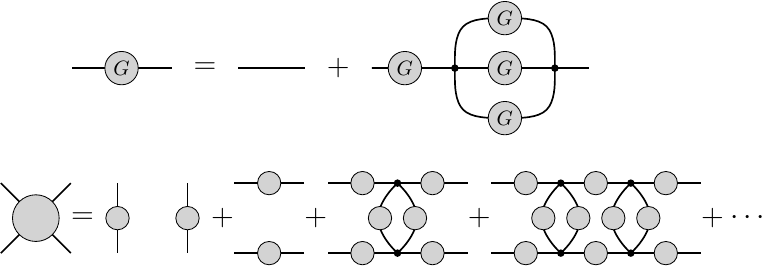}
    \caption{In the $N\to + \infty$ limit, the full two-point function of a melonic QFT obeys a closed Schwinger-Dyson equation (top), while the full four-point function is a sum of ladder diagrams (bottom).}
    \label{fig:sde}
\end{figure}

\begin{figure}
    \centering
    \includegraphics[scale=.8]{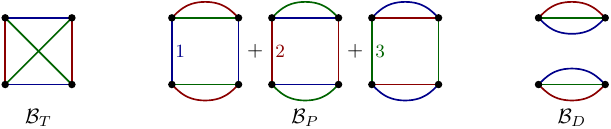}
    \caption{Quartic invariants of $\mathrm{O}(N)^3$ melonic QFTs.}
    \label{fig:On_quartic_invariants}
\end{figure}

It was first realized by Witten \cite{Witten:2016iux}, then by Klebanov and Tarnopolsky \cite{Klebanov:2016xxf}, that tensor models are able to emulate the physics of the Sachdev-Ye-Kitaev model \cite{Sachdev:1992fk, kitaev2015simple, Maldacena:2016hyu}. The latter describes a \emph{disordered} quantum-mechanical system of $N$ Majorana fermions $\psi_a$ ($a \in \{ 1, \ldots , N\}$), with all-to-all interaction Hamiltonian $H \sim J_{a b c d} \psi_a \psi_b \psi_c \psi_d$. Taking the random couplings $J_{a b c d}$ to be independent centered Gaussian variables with $N^3\langle J_{abcd}^2\rangle = \lambda^2$ leads to a consistent large $N$ limit (for the quenched theory). If one takes in addition the strong coupling limit $\lambda \gg 1$, an emergent time-reparametrization symmetry makes the model explicitly solvable. This symmetry and the way in which it is (spontaneously and explicitly) broken, leading to an effective Schwarzian dynamics \cite{Maldacena:2016hyu, Stanford:2017thb}, is structurally identical to what is found in a two-dimensional model of gravity known as Jackiw-Teitelboim gravity \cite{Maldacena:2016upp}. As first described by Kitaev \cite{kitaev2015simple}, this observation leads to a rather explicit holographic duality between SYK-like quantum mechanics and $\mathrm{AdS}_2$ quantum gravity, which has received a lot of attention in recent years, notably for its applications to the theory of quantum chaos and quantum black holes (see e.g. the review \cite{Mertens:2022irh}). Interestingly for us here, a key property of the SYK model is that its large $N$ limit is dominated by \emph{melonic} Feynman diagrams (see e.g. \cite{Bonzom:2017pqs}), which can be explicitly resummed in the strong coupling regime. To be more precise, this melonic limit is of a \emph{bilocal} type, which turns out to be essential to generate the interesting physics we just hinted at. The pioneering works of Witten and Klebanov-Tarnopolsky recognized that the bilocal melonic limits of tensor models can also support SYK-like physics, with the advantage that \emph{no disorder} needs to be introduced. Witten's model \cite{Witten:2016iux} was formulated in the original colored framework proposed by Gurau \cite{Gurau:2010ba}, while the Klebanov-Tarnopolsky model \cite{Klebanov:2016xxf} relied on the 'uncolored' combinatorial structure of the $\mathrm{O}(N)^3$ model \cite{Carrozza:2015adg}. Since the latter has imposed itself as the large $N$ structure of choice in later works, we focus on the Klebanov-Tarnopolsky model here. It describes $N^3$ Majorana fermions $\psi_{abc}$ ($a,b,c \in \{ 1, \ldots , N\}$), with action 
\begin{equation}\label{eq:KT_model}
S = \int dt \left( \frac{\mathrm{i}}{2} \psi_{abc}(t) \partial_t \psi_{abc}(t) + \frac{\lambda}{4 N^{3/2}} \psi_{abc}(t)\psi_{de c}(t) \psi_{d b f}(t)\psi_{a e f}(t)\right)\,.
\end{equation}
Note that the interaction part of the action is given by a 'tetrahedral' invariant as in \eqref{eq:action_sym} (corresponding to the colored bubble $\cB_T$ of Fig.~\ref{fig:On_quartic_invariants}), the main difference being the non-trivial time-dependence. We have in particular a non-trivial propagator, and a deterministic interaction which is local in time. The structure of the large $N$ limit is identical to the one described in \eqref{eq:largeN_On}, except that the Feynman amplitudes and fixed-degree partition functions are more complicated. In particular, in the $N \to + \infty$ we have a closed equation for the full two-point function \cite{Klebanov:2016xxf, kitaev2015simple, Maldacena:2016hyu}
\begin{equation}
    G(t_1 ,t_2) = G_{\mathrm{free}} (t_1 , t_2) + \lambda^2 \, \int \extd t \extd t' \, G_{\mathrm{free}} (t_1 , t)  \, [G(t,t')]^3 \, G(t',t_2)\,,
\end{equation}
where $G(t_1 , t_2 ) := N^{-3}\langle T( \psi_{a_1 a_2 a_3} (t_1 ) \psi_{a_1 a_2 a_3} (t_2 ))  \rangle$ and $G_{\rm free}(t_1 , t_2)$ denotes the same quantity at $\lambda = 0$. This equation is structurally similar to \eqref{eq:fuss-catalan}, and actually results from the same melonic Schwinger-Dyson equation, as represented in Fig.~\ref{fig:sde}. In the strong-coupling regime $\lambda \gg 1$, the second term on the right-hand side can be shown to dominate over the first one \cite{kitaev2015simple, Maldacena:2016hyu}, which leads to the simplified equation
\begin{equation}
\lambda^2 \, \int \extd t \, G(t_1 , t) \, [G (t,t_2)]^{3} = - \delta( t_1 - t_2)  \,. 
\end{equation}
Remarkably, this equation is invariant under reparametrizations of time
\begin{equation}\label{eq:melonic_KT}
    t \to f(t) \,, \qquad G(t_1 , t_2) \to \vert f'(t_1) f'(t_2)\vert^{1/4} G(f(t_1), f(t_2))\,,  
\end{equation}
and admits a conformal solution $G(t_1 , t_2) \propto \mathrm{sgn}(t_1 - t_2) \vert t_1 - t_2\vert^{- 2 \Delta} $ of conformal dimension $\Delta=1/4$. This solution spontaneously breaks the reparametrization symmetry down to $\mathrm{SL}(2, \mathbb{R})$, leading to Goldstone modes $f \in \mathrm{Diff}(S^1)/\mathrm{SL}(2, \mathbb{R})$.\footnote{We are assuming an $S^1$ topology for the (Euclidean) time direction, which is appropriate for a thermal path-integral.} The effective dynamics of $f$ can in turn be determined by analysing the explicit breaking of the reparametrization symmetry introduced by $G_{\rm free}(t_1 , t_2)$ in \eqref{eq:melonic_KT}. This can be done analytically thanks to another structurally important feature of bilocal melonic large $N$ expansions: their leading order four-point functions can be resummed explicitly as geometric series of ladder diagrams. This allows to compute the leading inverse coupling correction to the conformal dynamics, which leads to an effective dynamics for the Goldstone modes governed by the so-called \emph{Schwarzian action}
\begin{equation}
    S_{\rm eff} (f) \sim \frac{1}{\lambda}  \int \extd t \, \{ f , t\}\,, \qquad \{ f, t \} = \frac{f'''(t)}{f'(t)} - \frac{3}{2} \left( \frac{f''(t)}{f'(t)}\right)^2\,.
\end{equation}
The observation that 
the same Schwarzian dynamics
governs boundary modes of Jackiw-Teitelboim gravity on the hyperbolic disk \cite{Maldacena:2016upp}, has led to the development of the much studied holographic correspondence between SYK-like many-body models and quantum gravity on $\mathrm{AdS}_2$ \cite{Mertens:2022irh}. 


{}


An interesting feature of the Klebanov-Tarnopolsky model of \eqref{eq:KT_model} over the standard SYK model is that it naturally fits in the framework of local QFT, and can in particular be generalized to higher spacetime dimensions. This has been taken advantage of in recent years to construct a new family of large $N$ QFTs \cite{Klebanov:2018fzb, Gurau:2019qag, Benedetti:2020seh} which enjoy similarly nice analytical properties as the SYK model: namely, the full two-point function obeys a closed Schwinger-Dyson equation at leading order, while information about four-point functions (as well as higher-point correlators) can be recovered from a resummation of certain classes of ladder diagrams; see Fig.~\ref{fig:sde}. In particular, one can consider a real bosonic tensor field $\phi_{abc}(x)$ in $d$ spacetime dimensions, with $\mathrm{O}(N)^3$-invariant (and color symmetric) action 
\begin{align}\label{eq:melonic_QFT}
    S(\phi) &= \frac{1}{2} \int \extd^d x   \phi_{abc}(x) \left( (-\Delta)^\zeta + m^{2\zeta} \right) \phi_{abc}(x)   \\
     &+ \frac{1}{4} \int \extd^d x \left( \frac{\lambda}{N^{3/2}} \Tr_{\cB_T} (\phi(x))  + \frac{\lambda_P}{N^{2}} \Tr_{\cB_P} (\phi(x)) + \frac{\lambda_D}{N^{3}} \Tr_{\cB_D} (\phi(x)) \right)\,, \nn
\end{align}
where $\cB_{P}$, $\cB_{D}$ and $\cB_T$ are the quartic invariants represented in Fig.~\ref{fig:On_quartic_invariants}, and $\zeta \in ] 0 , 1 ]$ is some fixed parameter. $B_T$ is nothing but the 'tetrahedral' interaction, while $\cB_{P}$ ('pillow') and $\cB_{D}$ ('double-trace') need to be included for consistency of the renormalization group flow. This class of models was first considered in \cite{Giombi:2017dtl} with the standard $\left( -\Delta + m^2 \right)^{-1}$ propagator (that is, setting $\zeta = 1$ in \eqref{eq:melonic_QFT}) and in $d = 4 - \epsilon$. It was shown that, at leading order in $1/N$ and $\epsilon$, the $\beta$-functions of the model admit a non-trivial fixed point with $\lambda$, $\lambda_T$ and $\lambda_P$ all real of order $\sqrt{\epsilon}$. This defines an infrared conformal field theory, which is however unstable due to the presence of an operator of complex dimension \cite{Giombi:2017dtl, Benedetti:2021qyk}. This instability motivated a modification of the model \cite{Benedetti:2019eyl} in which: 1) the propagator is taken to be \emph{long-range}, meaning that $\zeta < 1$ in \eqref{eq:melonic_QFT}; 2) the 'tetrahedral' coupling $\lambda$ is assumed to be \emph{purely imaginary}. More precisely, setting $\zeta = d/4$ makes the quartic interaction marginal, and as a consequence of the long-range propagator, there is no wave-function renormalization. This simplification, together with the melonic structure, allows to compute flow equations non-perturbatively in the coupling constants, and to construct explicit renormalization group trajectories that flow to a unitary CFT in the IR \cite{Benedetti:2019ikb, Benedetti:2020yvb, Harribey:2022esw}. This result provides a well-controlled mathematical framework in which to test key paradigms of QFT: 
for instance, it was proven that the long-range model in $d=3$ obeys the $F$-theorem \cite{Benedetti:2021wzt}. Finally, the model \eqref{eq:melonic_QFT} with purely imaginary 'tetrahedral' coupling constant was also considered for $\zeta =1$ (short-range) and $d=4$, where it was observed that the $\beta$-functions are consistent with a well-behaved and \emph{asymptotically free} theory in the ultraviolet \cite{Berges:2023rqa}. Further results on tensor quantum mechanics and tensor QFT can be found in e.g.~\cite{Peng:2016mxj, Choudhury:2017tax, Bulycheva:2017ilt, Prakash:2017hwq, Benedetti:2017fmp, Benedetti:2018goh,  Benedetti:2018ghn, Klebanov:2018nfp, Giombi:2018qgp, Kim:2019upg, Pakrouski:2018jcc, Benedetti:2019rja,  Benedetti:2019sop, DeMelloKoch:2019tmo, Benedetti:2020iku, Benedetti:2020sye, Harribey:2021xgh, Jepsen:2023pzm}. 

\section{Group field theories}\label{sec:GFT}

\subsection{Nonlocal field theories of a tensorial type}

The framework of GFT \cite{Oriti:2006se, Freidel:2005qe, Krajewski:2011zzu} can be understood as a further generalization of tensor models, in which tensor 'indices' are taken to live in a Lie group $G$, which we will assume to be compact. This extra group structure can be exploited to equip the stranded graphs of tensor models with lattice gauge theory data, and weigh their amplitudes accordingly. In a similar way as the Feynman expansion of a tensor model can be interpreted as generating a random (discrete) geometry, the Feynman expansion of a GFT can be understood as defining a lattice gauge theory on a random lattice. Prototypical examples are provided by the Boulatov and Ooguri models \cite{Boulatov:1992vp, Ooguri:1992eb}. They are theories of a single field $\vphi \in L^2(G^D)$ obeying the so-called \emph{gauge invariance condition} (or \emph{closure constraint}):
\begin{equation}\label{eq:gauge_constraint}
    \forall h \in G\,, \qquad \vphi(g_1 h , g_2 h, \ldots , g_D h) = \vphi(g_1 , g_2 , \ldots , g_D)\,.
\end{equation}
In the template of equation \eqref{eq:measure}, this condition can be imposed by choosing $\pmb P$ to be the orthogonal projector on field configurations obeying the closure constraint, whose integral kernel is:
\begin{equation}\label{eq:propa_BO}
    \pmb P (g_1 , g_2 , \ldots , g_D ; g_1' , g_2' , \ldots , g_D') = \int_G \extd h \prod_{k = 1}^D \delta\left(g_k h g_k'^{\inv} \right)\,,
\end{equation}
and $h$ in this formula can ultimately be interpreted, on a given Feynman graph, as parallel transport between two interaction vertices connected by a propagator line.
The interaction originally considered by Boulatov \cite{Boulatov:1992vp}, with $D=3$ and $G=\SU(2)$, was a 'tetrahedral' interaction analogous to \eqref{eq:action_sym},\footnote{The Ooguri model is similarly defined in $D=4$, with an interaction following the combinatorial pattern of a $4$-simplex.} or in its colored version
\beq\label{eq:int_Boulatov}
S_{\rm int}(\vphi) = \frac{-\lambda}{4}  \int_{G^6} \vphi(g_1 , g_2 , g_3 ) \vphi( g_4 , g_5 , g_3) \vphi( g_4 , g_2 , g_6) \vphi( g_1 , g_5 , g_6)\,.
\eeq
Complex variants of this model, based on bipartite colored bubble interactions, started to be considered in the original papers on melonic large $N$ expansions \cite{Gurau:2009tw, Gurau:2010ba}, and made the structure of the random lattice better defined from a topological standpoint \cite{Gurau:2010nd, Gurau:2011xp}. In this framework, the Boulatov model generates $BF$ lattice gauge theory amplitudes which are topological invariants (modulo divergences that need to be regularized in an appropriate way). In the special case $G = \SU(2)$, those can be interpreted as three-dimensional Euclidean quantum gravity path-integrals expressed in holonomy-flux variables, and are mapped upon Fourier transform to the Ponzano-Regge state-sum model \cite{ponzano1968semiclassical}. This correspondence between the Boulatov and Ponzano-Regge models has motivated numerous further works exploring the relation between GFT and quantum gravity, and more specifically so in the context of loop quantum gravity and spin foam models \cite{DePietri:1999bx, Reisenberger:2000zc, BenGeloun:2010qkf, Baratin:2010wi, Baratin:2011tg, Baratin:2011hp,  Baratin:2011tx, Oriti:2013aqa, Jercher:2022mky}.\footnote{This is where the term \emph{group field theory} actually originates from.} From this perspective, GFT provides a tentative albeit natural completion of spin foam models,\footnote{See \cite{Asante:2022dnj} for an alternative approach.} one that enables the use of standard QFT concepts and methods (e.g. renormalization) to explore their phase structure. The most studied spin foam models of four-dimensional quantum gravity (see e.g. \cite{Perez:2012wv, Livine:2024hhc} for reviews), which share certain features of the simpler models we will restrict our attention to
, can all be embedded into the GFT framework.  


The GFT formalism aims at defining QFTs \emph{of} spacetime, rather than \emph{on} spacetime. It is therefore not constrained by the same notion of (spacetime) locality that is so fundamental in standard QFT. Indeed, from a mathematical standpoint, a GFT can be understood as a \emph{non-local} QFT on the manifold $G^D$, in which interactions obey a particular kind of non-locality (sometimes referred to as \emph{combinatorial non-locality}): for instance, the interaction \eqref{eq:int_Boulatov} involves an integration on $G^6$, which makes it non-local, but it has a rather specific structure in which individual group elements in $G$ interact locally by pairs. This pairwise locality is the essence of GFT; it constrains the kind of interactions one can legitimately consider in a similar way as the ordinary notion of spacetime locality does in standard QFT. 

\medskip

Like any other QFT, a GFT is typically plagued with divergences which arise from the short scale structure of the configuration space $G^D$. These must be regularized and renormalized away in some consistent manner. Two approaches have been considered to achieve this goal. In the first, one looks for an asymptotic large $N$ expansion analogous to the ones described in Sec.~\ref{sec:TM} for purely combinatorial models, where $N$ is identified to a cut-off within a suitable regularization scheme. For instance, one may consider a regularized version of the propagator \eqref{eq:propa_BO} 
\begin{equation}
    \pmb P_{1/N^2} (g_1 , g_2 , \ldots , g_D ; g_1' , g_2' , \ldots , g_D') = \int_{G} \extd h \prod_{k = 1}^D K_{1/N^2} \left(g_k h g_k'^{\inv} \right)\,,
\end{equation}
where $K_{1/N^2}$ is the heat-kernel at time $1/N^2$ on $G$, which imposes a smooth cut-off of order $N$ on dimensions of irreducible representations labelling harmonic modes.\footnote{E.g. when $G= \SU(2)$, we have
\begin{equation}
    K_\alpha (g) = \sum_{j \in \frac{\mathbb{N}}{2}} \exp\left( - \alpha j (j+1)\right) \chi_j (g)\,,
\end{equation}
where $\chi_j$ is the character of the spin-$j$ irreducible representation of $\SU(2)$.} This has the effect of regularizing any possible divergence of models of the Boulatov-Ooguri type, and one may then investigate which asymptotic scalings of their bubble interactions lead to consistent large $N$ limits. After pioneering steps were made in this direction \cite{Freidel:2009hd, Magnen:2009at, BenGeloun:2010wbk, BenGeloun:2010qkf, Krajewski:2010yq, Bonzom:2010ar, Bonzom:2010zh}
, the first large $N$ results of Gurau opened the way to a thorough investigation of complex and colored versions of the Boulatov-Ooguri models, which were found to obey a consistent large $N$ expansion dominated by melonic diagrams \cite{Gurau:2010ba, Gurau:2011aq, Gurau:2011xq, Bonzom:2011br, Carrozza:2011jn,  Carrozza:2012kt}. The critical properties of the leading-order sector have also been analyzed in some respects \cite{Baratin:2013rja}, but remain significantly less understood than for colored tensor models. In particular, the natural metric interpretation of GFT Feynman graphs is function of their combinatorial structure, but also of the group data attached to their edges. This makes a precise analysis of GFT models of random geometry difficult. To our knowledge, no definitive results have been obtain regarding the metric properties of their continuum limit. 

The second approach to a rigorous definition of GFTs embraces their interpretation as QFTs, and looks for an appropriate generalization of the full machinery of (perturbative and non-perturbative) renormalization. An important challenge is that no a priori notion of spacetime scale is available in GFT, and neither is the usual notion of locality, so it is not immediately clear what one would even mean by a flow of local interactions parametrized by a scale. This problem was first addressed and resolved in the context of GFTs in which no gauge invariance condition has been imposed, which are known in the literature as (\emph{tensor} or) \emph{tensorial field theories} \cite{BenGeloun:2011rc, Rivasseau:2011hm, BenGeloun:2012pu} (not to be confused with the local field theories of Sec.~\ref{sec:local_tft}!). Such models do not exploit the group structure to   
produce a sum over lattice gauge theory or quantum gravity amplitudes, and have in particular no direct connection to spin foams, but they remain mathematically interesting in their own right.






\subsection{Renormalization}\label{sec:GFT_ren}

In standard QFT, the renormalization group method relies on an harmonious interplay between spacetime scales and locality, in the sense that high energy radiative corrections to local interactions remain local as seen from low energy. In tensorial field theories, it was proposed that colored tensorial invariance should play the role of locality, meaning that, in complete analogy with tensor models, basic interactions should be labeled by connected colored graphs encoding pairwise convolutions of the group variables of the field. In an ordinary QFT, the energy scale ladder can be put in correspondence with the non-trivial spectrum of the propagator. In the GFT context, this motivates the introduction of a more abstract notion of scale, defined as an element of the spectrum of the propagator. Obviously, such a definition is relevant only in situations when the propagator has a rich spectrum, which is not the case of e.g. the Boulatov-Ooguri propagator \eqref{eq:propa_BO} (since it is a projector). This observation, together with other arguments \cite{BenGeloun:2011jnm}, motivated the introduction of tensorial field theories and GFTs with non-trivial propagators (such as e.g. $(-\Delta + m^2)^{-1}$ where $\Delta$ is the Laplace-Beltrami operator on $G^D$). 

This strategy was first proposed in the pioneering work \cite{BenGeloun:2011rc}, where it was proven that the abstract notions of locality and scale thus defined can support a consistent perturbative renormalization scheme for tensorial field theories. Moreover, a first all-order perturbative renormalizability theorem was obtained by means of rigorous multiscale renormalization methods \cite{rivasseau1991}. Further results of this kind were obtained in subsequent works, such as \cite{BenGeloun:2012pu, BenGeloun:2012yk, BenGeloun:2012phv, BenGeloun:2013vwi, BenGeloun:2017xbd, Geloun:2023oyd}. Interestingly, there is a close analogy between the perturbative renormalization analysis of tensorial field theories and the large $N$ analysis of tensor models. The \emph{superficial degree of divergence} of a Feynman graph $\cG$ can be expressed in terms of Gurau's degree \eqref{eq:degree}, while scaling dimensions of perturbative coupling constants are analogous to the scaling parameters $\alpha(\cB)$ from \eqref{eq:action_colored}. It is also crucial in the renormalization analysis that divergent graphs tend to have a simple combinatorial structure (such as a melonic one). Hence, from a qualitative point of view, it is possible to interpret the renormalization group flow of a tensorial field theory as dynamically generating a consistent asymptotic large $N$ scaling, similar to those that are being imposed by hand in the tensor model setting.   


The perturbative renormalization analysis of tensorial field theories was generalized in \cite{Carrozza:2012uv, Samary:2012bw, Carrozza:2013wda, carrozza2014tensorial} to GFT proper, that is to say to tensor field theories with gauge invariance condition \eqref{eq:gauge_constraint}. To be more precise, the class of GFT models considered in those works are complex models with regularized propagator\footnote{This is a regularized version of $(- \Delta + m^2)^{-1}$, in which the eigenvalues of $- \Delta + m^2$ of  order $\Lambda^2$ or larger have been smoothly cut-off.}
\begin{equation}
    \pmb P_{\Lambda} (g_1 , g_2 , \ldots , g_D ; g_1' , g_2' , \ldots , g_D') = \int_{1/ \Lambda^2}\extd \alpha \, e^{-\alpha m^2} \int_G \extd h \prod_{k = 1}^D K_\alpha \left(g_k h g_k'^{\inv} \right)\,, 
\end{equation}
and interactions labelled by connected $D$-colored diagrams. The \emph{superficial degree of divergence} of a Feynman graph $\cG$ has been determined to be of the form \cite{Carrozza:2013wda}
\begin{equation}\label{eq:divergence_degree}
    \omega(\cG) = - 2 L(\cG) + d_G \left( F(\cG) - R(\cG) \right) \,, 
\end{equation}
where $L(\cG)$ is the number of propagator lines of $\cG$, $F(\cG)$ its number of bi-colored cycles involving the color $0$ (or \emph{faces}), $R(\cG)$ is the rank of the adjacency matrix between its lines and faces, and $d_G$ the dimension of $G$. Divergences arise from subgraphs with $\omega \geq 0$, and formula \eqref{eq:divergence_degree} can be exploited to classify the set of potentially renormalizable models in terms of the free parameters $D$, $d_G$, as well as the list of bubble interactions included in the action \cite{Carrozza:2013wda, Carrozza:2016vsq}. And just like for models with no gauge invariance, the multiscale renormalization methods of \cite{rivasseau1991} turn out to be well-suited to the proof of all-order perturbative renormalizability theorems, because they allow to efficiently deal with the renormalization of overlapping divergent subgraphs. In particular, it was shown that with $D=3$ and $G= \SU(2)$, the sextic model involving the first four interactions shown in Fig.~\ref{fig:Un_tensors} is perturbatively renormalizable at all orders \cite{Carrozza:2013wda}, and its discrete multiscale flow equations were analyzed in detail in \cite{Carrozza:2014rba}. The fact that ultraviolet divergences can be reabsorbed into effective interactions which still obey the criterion of tensorial locality is non-trivial in this model; it relies on a nice interplay between the topology of the Feynman graphs (as defined via Pezzana's duality) and the lattice gauge theory structure of their amplitudes. We refer to the review \cite{Carrozza:2016vsq} and references therein for more details on this class of models. 

Other interesting results have been obtained about the renormalization of tensorial field theories and GFTs. To begin with, it was established in \cite{BenGeloun:2012yk} that tensor field theories can generate renormalization group flows which are \emph{asymptotically free} in the utraviolet. This may appear surprising given their superficial similarity to standard scalar field theories (which are not compatible with asymptotic freedom), but it turns out to be a consequence of the non-trivial combinatorial structure of tensorial interactions. In \cite{Rivasseau:2015ova}, it was shown that wave-function renormalization tends to dominate over vertex renormalization, and that, as a result, asymptotic freedom is a robust feature of quartic models. The situation is more complicated in models involving marginal interactions of higher order, as exemplified by the sextic model of \cite{Carrozza:2013wda}, which fails to be asymptotically free for somewhat subtle reasons \cite{Rivasseau:2015ova}. Nonetheless, it became clear from these results that tensor field theory provides non-trivial examples of QFTs which are mathematically interesting in their own right, independently from their eventual relation to GFT or quantum gravity. Since then, tensor field theory has been used as a test bed for constructive renormalization programs: for instance, the multiscale renormalization methods of \cite{rivasseau1991, Gurau:2013oqa, Rivasseau:2017hpg} have been employed to construct renormalizable tensorial field theories in \cite{Delepouve:2014hfa, Rivasseau:2017xbk, Rivasseau:2021rlt}, and more recently, such models were revisited through the lenses of stochastic analysis \cite{Chandra:2023kpp}. In another direction, the perturbative renormalization struture underlying tensor field theories was shown to fit in the general Hopf-algebraic framework due to Connes and Kreimer \cite{Raasakka:2013kaa, Avohou:2015sia, Thurigen:2021zwf, Thurigen:2021ntr}. Finally, a number of efforts have been made to map out the non-perturbative fixed points of tensor field theories and GFTs by means of Functional Renormalization Group (FRG) methods. This program started out with the analysis of \cite{Benedetti:2014qsa, BenGeloun:2015xrk}, which focused on models without gauge invariance condition, and was then applied to proper GFTs in \cite{Carrozza:2014rya, Benedetti:2015yaa, BenGeloun:2016rqa, Carrozza:2016tih, Carrozza:2017vkz, Lahoche:2016xiq}. As far as quantum gravity is concerned, it is hoped that non-perturbative FRG methods in appropriate truncations will help map out the phase diagrams of GFTs in the infrared regime, and eventually allow to locate non-perturbative fixed points which have the potential to lead to interesting continuum limits. We refer to the recent literature \cite{BenGeloun:2018ekd, Pithis:2020sxm, Pithis:2020kio, Geloun:2023ray} for a detailed account and status of this search (see also \cite{Eichhorn:2017xhy, Eichhorn:2018phj} for applications of functional renormalization to the search of consistent large $N$ limits). 

We conclude this section with two open questions regarding the renormalization of GFTs. First, while rigorous perturbative renormalization theorems are already available for simple classes of GFTs (tensor field theories, and GFTs resembling the Boulatov-Ooguri models), can those results and the methods they rest on be extended to four-dimensional quantum gravity proposals (e.g. of the type \cite{Baratin:2011hp, Baratin:2011tx, Oriti:2013aqa, Finocchiaro:2020fhl, Jercher:2022mky})? Second, how could one go about proving the existence of the non-perturbative fixed points which are empirically observed in truncated FRGs? 





\subsection{Further developments}\label{subsec:GFT-cosmo}

We conclude by listing some important research directions on GFT which we could not give justice to in this short review. 

In recent years, there has been a strong push towards applications of the GFT formalism to cosmology, and other symmetry-reduced regimes of general relativity. A mechanism of Bose-Einstein condensation adapted to GFT was investigated and argued to predict effective Friedmann equations from full-fledged quantum gravity \cite{Gielen:2013kla, Gielen:2013naa, Oriti:2016qtz}. A similar construction was also applied to black holes in \cite{Oriti:2015rwa, Oriti:2018qty}. This approach is grounded in the Fock space formalism of \cite{Oriti:2013aqa}; it allows to extract a low energy effective dynamics from GFT by means of well-established methods originating from standard quantum many-body physics. We refer to the reviews \cite{Gielen:2016dss, Oriti:2016acw, Pithis:2019tvp} for more complete accounts of GFT cosmology.

The same many-body physics interpretation has also allowed to analyze GFT wave-functions through the lenses of entanglement theory. In particular, holographic entanglement area laws \cite{Chirco:2017vhs, Chirco:2017wgl, Chirco:2019gsd} have been derived in GFT, thus establishing interesting connections with holographic tensor networks \cite{Hayden:2016cfa, Colafranceschi:2020ern}. 

While the GFTs we focused on in the previous two subsections assumed a \emph{compact} Lie group $G$, GFT models of \emph{Lorentzian} quantum gravity typically require $G$ to be \emph{non-compact} (see e.g. \cite{Jercher:2021bie, Jercher:2022mky}). From the point of view of renormalization, this leads to an additional infrared problem, which was approached by means of FRG methods in \cite{BenGeloun:2015xrk, BenGeloun:2016rqa}. Interestingly, in recent works \cite{Marchetti:2022igl, Marchetti:2022nrf}, the non-compact nature of $G$ was taken advantage of to develop new mean-field approximations, which provide useful information about those models while a complete renormalization group treatment awaits.

Finally, we note that a certain type of effective melonic quantum dynamics was recently suggested to arise from a generalized Boulatov model \cite{Nador:2023inw}. This may lead to interesting new connections between GFT and the large $N$ QFT framework described in Sec.~\ref{sec:local_tft}, which are a priori quite different (both in intent and content). 



\section*{Acknowledgements}

I would like to thank Razvan Gurau, Daniele Oriti and Vincent Rivasseau for their pioneering works, their contagious enthusiasm for this subject, and for teaching me much about it in the first place. 

\medskip

\noindent The IMB receives support from the EIPHI Graduate School (contract ANR-17-EURE-0002).

{\footnotesize
\bibliographystyle{JHEP}
\addcontentsline{toc}{section}{References}
\let\oldbibliography\thebibliography
\renewcommand{\thebibliography}[1]{\oldbibliography{#1}\setlength{\itemsep}{-1pt}}
\bibliography{biblio} }

\providecommand{\href}[2]{#2}\begingroup\raggedright\begin{thebibliography}{100}

\bibitem{tHooft:1973alw}
G.~'t~Hooft, \emph{{A Planar Diagram Theory for Strong Interactions}},
  \href{https://doi.org/10.1016/0550-3213(74)90154-0}{\emph{Nucl. Phys. B}
  {\bfseries 72} (1974) 461}.

\bibitem{lando2004graphs}
S.~K. Lando, A.~K. Zvonkin and D.~B. Zagier, \emph{Graphs on surfaces and their
  applications}, vol.~141. Springer, 2004.

\bibitem{Brezin:1977sv}
E.~Brezin, C.~Itzykson, G.~Parisi and J.~B. Zuber, \emph{{Planar Diagrams}},
  \href{https://doi.org/10.1007/BF01614153}{\emph{Commun. Math. Phys.}
  {\bfseries 59} (1978) 35}.

\bibitem{David:1984tx}
F.~David, \emph{{Planar Diagrams, Two-Dimensional Lattice Gravity and Surface
  Models}}, \href{https://doi.org/10.1016/0550-3213(85)90335-9}{\emph{Nucl.
  Phys. B} {\bfseries 257} (1985) 45}.

\bibitem{Ambjorn:1985az}
J.~Ambjorn, B.~Durhuus and J.~Frohlich, \emph{{Diseases of Triangulated Random
  Surface Models, and Possible Cures}},
  \href{https://doi.org/10.1016/0550-3213(85)90356-6}{\emph{Nucl. Phys. B}
  {\bfseries 257} (1985) 433}.

\bibitem{DiFrancesco:1993cyw}
P.~Di~Francesco, P.~H. Ginsparg and J.~Zinn-Justin, \emph{{2-D Gravity and
  random matrices}},
  \href{https://doi.org/10.1016/0370-1573(94)00084-G}{\emph{Phys. Rept.}
  {\bfseries 254} (1995) 1}
  [\href{https://arxiv.org/abs/hep-th/9306153}{{\ttfamily hep-th/9306153}}].

\bibitem{Eynard:2015aea}
B.~Eynard, T.~Kimura and S.~Ribault, \emph{{Random matrices}},
  \href{https://arxiv.org/abs/1510.04430}{{\ttfamily 1510.04430}}.

\bibitem{Eynard:2016yaa}
B.~Eynard, \emph{{Counting Surfaces}}, vol.~70 of \emph{Progress in
  Mathematical Physics}. Springer, 2016,
  \href{https://doi.org/10.1007/978-3-7643-8797-6}{10.1007/978-3-7643-8797-6}.

\bibitem{le2013uniqueness}
J.-F. Le~Gall, \emph{Uniqueness and universality of the brownian map},
  {\emph{The Annals of Probability} {\bfseries 41} (2013) 2880}.

\bibitem{miermont2013brownian}
G.~Miermont, \emph{The brownian map is the scaling limit of uniform random
  plane quadrangulations}, {\emph{Acta Math} {\bfseries 210} (2013) 319}.

\bibitem{miller2015liouville}
J.~Miller and S.~Sheffield, \emph{Liouville quantum gravity and the brownian
  map i: The qle (8/3, 0) metric}, {\emph{arXiv preprint arXiv:1507.00719}
  (2015) }.

\bibitem{gwynne2020random}
E.~Gwynne, \emph{Random surfaces and liouville quantum gravity}, {\emph{Notices
  of the American Mathematical Society} {\bfseries 67} (2020) 484}.

\bibitem{Ambjorn:1990ge}
J.~Ambjorn, B.~Durhuus and T.~Jonsson, \emph{{Three-dimensional simplicial
  quantum gravity and generalized matrix models}},
  \href{https://doi.org/10.1142/S0217732391001184}{\emph{Mod. Phys. Lett. A}
  {\bfseries 6} (1991) 1133}.

\bibitem{Sasakura:1990fs}
N.~Sasakura, \emph{{Tensor model for gravity and orientability of manifold}},
  \href{https://doi.org/10.1142/S0217732391003055}{\emph{Mod. Phys. Lett. A}
  {\bfseries 6} (1991) 2613}.

\bibitem{Gross:1991hx}
M.~Gross, \emph{{Tensor models and simplicial quantum gravity in \ensuremath{>}
  2-D}}, \href{https://doi.org/10.1016/S0920-5632(05)80015-5}{\emph{Nucl. Phys.
  B Proc. Suppl.} {\bfseries 25} (1992) 144}.

\bibitem{Oriti:2006se}
D.~Oriti, \emph{{The Group field theory approach to quantum gravity}},
  \href{https://arxiv.org/abs/gr-qc/0607032}{{\ttfamily gr-qc/0607032}}.

\bibitem{Freidel:2005qe}
L.~Freidel, \emph{{Group field theory: An Overview}},
  \href{https://doi.org/10.1007/s10773-005-8894-1}{\emph{Int. J. Theor. Phys.}
  {\bfseries 44} (2005) 1769}
  [\href{https://arxiv.org/abs/hep-th/0505016}{{\ttfamily hep-th/0505016}}].

\bibitem{Krajewski:2011zzu}
T.~Krajewski, \emph{{Group field theories}},
  \href{https://doi.org/10.22323/1.140.0005}{\emph{PoS} {\bfseries QGQGS2011}
  (2011) 005} [\href{https://arxiv.org/abs/1210.6257}{{\ttfamily 1210.6257}}].

\bibitem{Boulatov:1992vp}
D.~V. Boulatov, \emph{{A Model of three-dimensional lattice gravity}},
  \href{https://doi.org/10.1142/S0217732392001324}{\emph{Mod. Phys. Lett. A}
  {\bfseries 7} (1992) 1629}
  [\href{https://arxiv.org/abs/hep-th/9202074}{{\ttfamily hep-th/9202074}}].

\bibitem{Ooguri:1992eb}
H.~Ooguri, \emph{{Topological lattice models in four-dimensions}},
  \href{https://doi.org/10.1142/S0217732392004171}{\emph{Mod. Phys. Lett. A}
  {\bfseries 7} (1992) 2799}
  [\href{https://arxiv.org/abs/hep-th/9205090}{{\ttfamily hep-th/9205090}}].

\bibitem{DePietri:1999bx}
R.~De~Pietri, L.~Freidel, K.~Krasnov and C.~Rovelli, \emph{{Barrett-Crane model
  from a Boulatov-Ooguri field theory over a homogeneous space}},
  \href{https://doi.org/10.1016/S0550-3213(00)00005-5}{\emph{Nucl. Phys. B}
  {\bfseries 574} (2000) 785}
  [\href{https://arxiv.org/abs/hep-th/9907154}{{\ttfamily hep-th/9907154}}].

\bibitem{Reisenberger:2000zc}
M.~P. Reisenberger and C.~Rovelli, \emph{{Space-time as a Feynman diagram: The
  Connection formulation}},
  \href{https://doi.org/10.1088/0264-9381/18/1/308}{\emph{Class. Quant. Grav.}
  {\bfseries 18} (2001) 121}
  [\href{https://arxiv.org/abs/gr-qc/0002095}{{\ttfamily gr-qc/0002095}}].

\bibitem{Baratin:2011hp}
A.~Baratin and D.~Oriti, \emph{{Group field theory and simplicial gravity path
  integrals: A model for Holst-Plebanski gravity}},
  \href{https://doi.org/10.1103/PhysRevD.85.044003}{\emph{Phys. Rev. D}
  {\bfseries 85} (2012) 044003}
  [\href{https://arxiv.org/abs/1111.5842}{{\ttfamily 1111.5842}}].

\bibitem{Baratin:2011tx}
A.~Baratin and D.~Oriti, \emph{{Quantum simplicial geometry in the group field
  theory formalism: reconsidering the Barrett-Crane model}},
  \href{https://doi.org/10.1088/1367-2630/13/12/125011}{\emph{New J. Phys.}
  {\bfseries 13} (2011) 125011}
  [\href{https://arxiv.org/abs/1108.1178}{{\ttfamily 1108.1178}}].

\bibitem{Gurau:2009tw}
R.~Gurau, \emph{{Colored Group Field Theory}},
  \href{https://doi.org/10.1007/s00220-011-1226-9}{\emph{Commun. Math. Phys.}
  {\bfseries 304} (2011) 69} [\href{https://arxiv.org/abs/0907.2582}{{\ttfamily
  0907.2582}}].

\bibitem{Gurau:2010ba}
R.~Gurau, \emph{{The 1/N expansion of colored tensor models}},
  \href{https://doi.org/10.1007/s00023-011-0101-8}{\emph{Annales Henri
  Poincare} {\bfseries 12} (2011) 829}
  [\href{https://arxiv.org/abs/1011.2726}{{\ttfamily 1011.2726}}].

\bibitem{Gurau:2011aq}
R.~Gurau and V.~Rivasseau, \emph{{The 1/N expansion of colored tensor models in
  arbitrary dimension}},
  \href{https://doi.org/10.1209/0295-5075/95/50004}{\emph{EPL} {\bfseries 95}
  (2011) 50004} [\href{https://arxiv.org/abs/1101.4182}{{\ttfamily
  1101.4182}}].

\bibitem{Gurau:2011xq}
R.~Gurau, \emph{{The complete 1/N expansion of colored tensor models in
  arbitrary dimension}},
  \href{https://doi.org/10.1007/s00023-011-0118-z}{\emph{Annales Henri
  Poincare} {\bfseries 13} (2012) 399}
  [\href{https://arxiv.org/abs/1102.5759}{{\ttfamily 1102.5759}}].

\bibitem{Gurau:2024nzv}
R.~Gurau and V.~Rivasseau, \emph{{Quantum Gravity and Random Tensors}},
  \href{https://arxiv.org/abs/2401.13510}{{\ttfamily 2401.13510}}.

\bibitem{Gurau:2011kk}
R.~Gurau, \emph{{Universality for Random Tensors}},
  \href{https://doi.org/10.1214/13-AIHP567}{\emph{Ann. Inst. H. Poincare
  Probab. Statist.} {\bfseries 50} (2014) 1474}
  [\href{https://arxiv.org/abs/1111.0519}{{\ttfamily 1111.0519}}].

\bibitem{Gurau:2013pca}
R.~Gurau, \emph{{The 1/N Expansion of Tensor Models Beyond Perturbation
  Theory}}, \href{https://doi.org/10.1007/s00220-014-1907-2}{\emph{Commun.
  Math. Phys.} {\bfseries 330} (2014) 973}
  [\href{https://arxiv.org/abs/1304.2666}{{\ttfamily 1304.2666}}].

\bibitem{BenGeloun:2013lim}
J.~Ben~Geloun and S.~Ramgoolam, \emph{{Counting tensor model observables and
  branched covers of the 2-sphere}},
  \href{https://doi.org/10.4171/aihpd/4}{\emph{Ann. Inst. H. Poincare D Comb.
  Phys. Interact.} {\bfseries 1} (2014) 77}
  [\href{https://arxiv.org/abs/1307.6490}{{\ttfamily 1307.6490}}].

\bibitem{BenGeloun:2020lfe}
J.~Ben~Geloun, \emph{{On the counting tensor model observables as $U(N)$ and
  $O(N)$ classical invariants}},
  \href{https://doi.org/10.22323/1.376.0175}{\emph{PoS} {\bfseries CORFU2019}
  (2020) 175} [\href{https://arxiv.org/abs/2005.01773}{{\ttfamily
  2005.01773}}].

\bibitem{Gurau:2010nd}
R.~Gurau, \emph{{Lost in Translation: Topological Singularities in Group Field
  Theory}}, \href{https://doi.org/10.1088/0264-9381/27/23/235023}{\emph{Class.
  Quant. Grav.} {\bfseries 27} (2010) 235023}
  [\href{https://arxiv.org/abs/1006.0714}{{\ttfamily 1006.0714}}].

\bibitem{Bonzom:2011zz}
V.~Bonzom, R.~Gurau, A.~Riello and V.~Rivasseau, \emph{{Critical behavior of
  colored tensor models in the large N limit}},
  \href{https://doi.org/10.1016/j.nuclphysb.2011.07.022}{\emph{Nucl. Phys. B}
  {\bfseries 853} (2011) 174}
  [\href{https://arxiv.org/abs/1105.3122}{{\ttfamily 1105.3122}}].

\bibitem{Gurau:2011xp}
R.~Gurau and J.~P. Ryan, \emph{{Colored Tensor Models - a review}},
  \href{https://doi.org/10.3842/SIGMA.2012.020}{\emph{SIGMA} {\bfseries 8}
  (2012) 020} [\href{https://arxiv.org/abs/1109.4812}{{\ttfamily 1109.4812}}].

\bibitem{Bonzom:2012hw}
V.~Bonzom, R.~Gurau and V.~Rivasseau, \emph{{Random tensor models in the large
  N limit: Uncoloring the colored tensor models}},
  \href{https://doi.org/10.1103/PhysRevD.85.084037}{\emph{Phys. Rev. D}
  {\bfseries 85} (2012) 084037}
  [\href{https://arxiv.org/abs/1202.3637}{{\ttfamily 1202.3637}}].

\bibitem{gurau2017random}
R.~Gurau, \emph{Random Tensors}. Oxford University Press, 2017.

\bibitem{ferri1986graph}
M.~Ferri, C.~Gagliardi and L.~Grasselli, \emph{A graph-theoretical
  representation of pl-manifolds—a survey on crystallizations},
  {\emph{Aequationes mathematicae} {\bfseries 31} (1986) 121}.

\bibitem{lionni2018colored}
L.~Lionni, \emph{Colored discrete spaces: higher dimensional combinatorial maps
  and quantum gravity}. Springer, 2018.

\bibitem{BenGeloun:2010wbk}
J.~Ben~Geloun, T.~Krajewski, J.~Magnen and V.~Rivasseau, \emph{{Linearized
  Group Field Theory and Power Counting Theorems}},
  \href{https://doi.org/10.1088/0264-9381/27/15/155012}{\emph{Class. Quant.
  Grav.} {\bfseries 27} (2010) 155012}
  [\href{https://arxiv.org/abs/1002.3592}{{\ttfamily 1002.3592}}].

\bibitem{Tanasa:2011ur}
A.~Tanasa, \emph{{Multi-orientable Group Field Theory}},
  \href{https://doi.org/10.1088/1751-8113/45/16/165401}{\emph{J. Phys. A}
  {\bfseries 45} (2012) 165401}
  [\href{https://arxiv.org/abs/1109.0694}{{\ttfamily 1109.0694}}].

\bibitem{Dartois:2013he}
S.~Dartois, V.~Rivasseau and A.~Tanasa, \emph{{The $1/N$ expansion of
  multi-orientable random tensor models}},
  \href{https://doi.org/10.1007/s00023-013-0262-8}{\emph{Annales Henri
  Poincare} {\bfseries 15} (2014) 965}
  [\href{https://arxiv.org/abs/1301.1535}{{\ttfamily 1301.1535}}].

\bibitem{Raasakka:2013eda}
M.~Raasakka and A.~Tanasa, \emph{{Next-to-leading order in the large $N$
  expansion of the multi-orientable random tensor model}},
  \href{https://doi.org/10.1007/s00023-014-0336-2}{\emph{Annales Henri
  Poincare} {\bfseries 16} (2015) 1267}
  [\href{https://arxiv.org/abs/1310.3132}{{\ttfamily 1310.3132}}].

\bibitem{Fusy:2014rba}
E.~Fusy and A.~Tanasa, \emph{{Asymptotic expansion of the multi-orientable
  random tensor model}},  \href{https://arxiv.org/abs/1408.5725}{{\ttfamily
  1408.5725}}.

\bibitem{Tanasa:2015uhr}
A.~Tanasa, \emph{{The Multi-Orientable Random Tensor Model, a Review}},
  \href{https://doi.org/10.3842/SIGMA.2016.056}{\emph{SIGMA} {\bfseries 12}
  (2016) 056} [\href{https://arxiv.org/abs/1512.02087}{{\ttfamily
  1512.02087}}].

\bibitem{Carrozza:2015adg}
S.~Carrozza and A.~Tanasa, \emph{{$O(N)$ Random Tensor Models}},
  \href{https://doi.org/10.1007/s11005-016-0879-x}{\emph{Lett. Math. Phys.}
  {\bfseries 106} (2016) 1531}
  [\href{https://arxiv.org/abs/1512.06718}{{\ttfamily 1512.06718}}].

\bibitem{Ferrari:2017jgw}
F.~Ferrari, V.~Rivasseau and G.~Valette, \emph{{A New Large $N$ Expansion for
  General Matrix\textendash{}Tensor Models}},
  \href{https://doi.org/10.1007/s00220-019-03511-7}{\emph{Commun. Math. Phys.}
  {\bfseries 370} (2019) 403}
  [\href{https://arxiv.org/abs/1709.07366}{{\ttfamily 1709.07366}}].

\bibitem{Bonzom:2019moj}
V.~Bonzom, \emph{{Another Proof of the $1/N$ Expansion of the Rank Three Tensor
  Model with Tetrahedral Interaction}},
  \href{https://arxiv.org/abs/1912.11104}{{\ttfamily 1912.11104}}.

\bibitem{Valette:2019nzp}
G.~Valette, \emph{{New Limits for Large N Matrix and Tensor Models: Large D,
  Melons and Applications}}, Ph.D. thesis, U. Brussels, U. Brussels (main),
  2019.
\newblock \href{https://arxiv.org/abs/1911.11574}{{\ttfamily 1911.11574}}.

\bibitem{Ferrari:2017ryl}
F.~Ferrari, \emph{{The large $D$ limit of planar diagrams}},
  \href{https://doi.org/10.4171/aihpd/76}{\emph{Ann. Inst. H. Poincare D Comb.
  Phys. Interact.} {\bfseries 6} (2019) 427}
  [\href{https://arxiv.org/abs/1701.01171}{{\ttfamily 1701.01171}}].

\bibitem{Azeyanagi:2017mre}
T.~Azeyanagi, F.~Ferrari, P.~Gregori, L.~Leduc and G.~Valette, \emph{{More on
  the New Large $D$ Limit of Matrix Models}},
  \href{https://doi.org/10.1016/j.aop.2018.04.010}{\emph{Annals Phys.}
  {\bfseries 393} (2018) 308}
  [\href{https://arxiv.org/abs/1710.07263}{{\ttfamily 1710.07263}}].

\bibitem{Benedetti:2020iyz}
D.~Benedetti, S.~Carrozza, R.~Toriumi and G.~Valette, \emph{{Multiple scaling
  limits of $\operatorname{U}(N)^2 \times \operatorname{O}(D)$ multi-matrix
  models}}, \href{https://doi.org/10.4171/aihpd/121}{\emph{Ann. Inst. H.
  Poincare D Comb. Phys. Interact.} {\bfseries 9} (2022) 367}
  [\href{https://arxiv.org/abs/2003.02100}{{\ttfamily 2003.02100}}].

\bibitem{Bonzom:2022yvc}
V.~Bonzom, V.~Nador and A.~Tanasa, \emph{{Double scaling limit of multi-matrix
  models at large D}}, \href{https://doi.org/10.1088/1751-8121/acb6c7}{\emph{J.
  Phys. A} {\bfseries 56} (2023) 075201}
  [\href{https://arxiv.org/abs/2209.02026}{{\ttfamily 2209.02026}}].

\bibitem{Avohou:2023wvg}
R.~C. Avohou, R.~Toriumi and M.~Vancraeynest, \emph{{Classification of higher
  grade $\ell$ graphs for $\mathrm{U}(N)^2\times \mathrm{O}(D)$ multi-matrix
  models}},  \href{https://arxiv.org/abs/2310.13789}{{\ttfamily 2310.13789}}.

\bibitem{Klebanov:2017nlk}
I.~R. Klebanov and G.~Tarnopolsky, \emph{{On Large $N$ Limit of Symmetric
  Traceless Tensor Models}},
  \href{https://doi.org/10.1007/JHEP10(2017)037}{\emph{JHEP} {\bfseries 10}
  (2017) 037} [\href{https://arxiv.org/abs/1706.00839}{{\ttfamily
  1706.00839}}].

\bibitem{Benedetti:2017qxl}
D.~Benedetti, S.~Carrozza, R.~Gurau and M.~Kolanowski, \emph{{The $1/N$
  expansion of the symmetric traceless and the antisymmetric tensor models in
  rank three}}, \href{https://doi.org/10.1007/s00220-019-03551-z}{\emph{Commun.
  Math. Phys.} {\bfseries 371} (2019) 55}
  [\href{https://arxiv.org/abs/1712.00249}{{\ttfamily 1712.00249}}].

\bibitem{Gurau:2017qya}
R.~Gurau, \emph{{The $1/N$ expansion of tensor models with two symmetric
  tensors}}, \href{https://doi.org/10.1007/s00220-017-3055-y}{\emph{Commun.
  Math. Phys.} {\bfseries 360} (2018) 985}
  [\href{https://arxiv.org/abs/1706.05328}{{\ttfamily 1706.05328}}].

\bibitem{Carrozza:2018ewt}
S.~Carrozza, \emph{{Large $N$ limit of irreducible tensor models: $O(N)$
  rank-$3$ tensors with mixed permutation symmetry}},
  \href{https://doi.org/10.1007/JHEP06(2018)039}{\emph{JHEP} {\bfseries 06}
  (2018) 039} [\href{https://arxiv.org/abs/1803.02496}{{\ttfamily
  1803.02496}}].

\bibitem{Carrozza:2021qos}
S.~Carrozza and S.~Harribey, \emph{{Melonic Large $N$ Limit of $5$-Index
  Irreducible Random Tensors}},
  \href{https://doi.org/10.1007/s00220-021-04299-1}{\emph{Commun. Math. Phys.}
  {\bfseries 390} (2022) 1219}
  [\href{https://arxiv.org/abs/2104.03665}{{\ttfamily 2104.03665}}].

\bibitem{Carrozza:2020eaz}
S.~Carrozza, F.~Ferrari, A.~Tanasa and G.~Valette, \emph{{On the large $D$
  expansion of Hermitian multi-matrix models}},
  \href{https://doi.org/10.1063/5.0008349}{\emph{J. Math. Phys.} {\bfseries 61}
  (2020) 073501} [\href{https://arxiv.org/abs/2003.04152}{{\ttfamily
  2003.04152}}].

\bibitem{Carrozza:2018psc}
S.~Carrozza and V.~Pozsgay, \emph{{SYK-like tensor quantum mechanics with
  $\mathrm{Sp}(N)$ symmetry}},
  \href{https://doi.org/10.1016/j.nuclphysb.2019.02.012}{\emph{Nucl. Phys. B}
  {\bfseries 941} (2019) 28}
  [\href{https://arxiv.org/abs/1809.07753}{{\ttfamily 1809.07753}}].

\bibitem{Gurau:2022dbx}
R.~Gurau and H.~Keppler, \emph{{Duality of Orthogonal and Symplectic Random
  Tensor Models}},  \href{https://arxiv.org/abs/2207.01993}{{\ttfamily
  2207.01993}}.

\bibitem{Keppler:2023lkb}
H.~Keppler and T.~Muller, \emph{{Duality of orthogonal and symplectic random
  tensor models: general invariants}},
  \href{https://doi.org/10.1007/s11005-023-01706-7}{\emph{Lett. Math. Phys.}
  {\bfseries 113} (2023) 83}
  [\href{https://arxiv.org/abs/2304.03625}{{\ttfamily 2304.03625}}].

\bibitem{carrance2021convergence}
A.~Carrance, \emph{Convergence of eulerian triangulations}, {\emph{Electronic
  Journal of Probability} {\bfseries 26} (2021) }.

\bibitem{flajolet2009analytic}
P.~Flajolet and R.~Sedgewick, \emph{Analytic combinatorics}. Cambridge
  University Press, 2009.

\bibitem{10.1214/aop/1176990534}
D.~Aldous, \emph{{The Continuum Random Tree. I}},
  \href{https://doi.org/10.1214/aop/1176990534}{\emph{The Annals of
  Probability} {\bfseries 19} (1991) 1 }.

\bibitem{Aldous_1991}
D.~Aldous, \emph{The continuum random tree ii: an overview},  in
  \emph{Stochastic Analysis: Proceedings of the Durham Symposium on Stochastic
  Analysis, 1990}, M.~T. Barlow and N.~H. Bingham, eds., London Mathematical
  Society Lecture Note Series, p.~23–70, Cambridge University Press, (1991).

\bibitem{Ambjorn:1990wp}
J.~Ambjorn, B.~Durhuus and T.~Jonsson, \emph{{Summing Over All Genera for $d$
  \ensuremath{>} 1: A Toy Model}},
  \href{https://doi.org/10.1016/0370-2693(90)90337-6}{\emph{Phys. Lett. B}
  {\bfseries 244} (1990) 403}.

\bibitem{Bialas:1996ya}
P.~Bialas and Z.~Burda, \emph{{Phase transition in fluctuating branched
  geometry}}, \href{https://doi.org/10.1016/0370-2693(96)00795-2}{\emph{Phys.
  Lett. B} {\bfseries 384} (1996) 75}
  [\href{https://arxiv.org/abs/hep-lat/9605020}{{\ttfamily hep-lat/9605020}}].

\bibitem{Gurau:2013cbh}
R.~Gurau and J.~P. Ryan, \emph{{Melons are branched polymers}},
  \href{https://doi.org/10.1007/s00023-013-0291-3}{\emph{Annales Henri
  Poincare} {\bfseries 15} (2014) 2085}
  [\href{https://arxiv.org/abs/1302.4386}{{\ttfamily 1302.4386}}].

\bibitem{Kaminski:2013maa}
W.~Kami\'nski, D.~Oriti and J.~P. Ryan, \emph{{Towards a double-scaling limit
  for tensor models: probing sub-dominant orders}},
  \href{https://doi.org/10.1088/1367-2630/16/6/063048}{\emph{New J. Phys.}
  {\bfseries 16} (2014) 063048}
  [\href{https://arxiv.org/abs/1304.6934}{{\ttfamily 1304.6934}}].

\bibitem{Dartois:2013sra}
S.~Dartois, R.~Gurau and V.~Rivasseau, \emph{{Double Scaling in Tensor Models
  with a Quartic Interaction}},
  \href{https://doi.org/10.1007/JHEP09(2013)088}{\emph{JHEP} {\bfseries 09}
  (2013) 088} [\href{https://arxiv.org/abs/1307.5281}{{\ttfamily 1307.5281}}].

\bibitem{Bonzom:2014oua}
V.~Bonzom, R.~Gurau, J.~P. Ryan and A.~Tanasa, \emph{{The double scaling limit
  of random tensor models}},
  \href{https://doi.org/10.1007/JHEP09(2014)051}{\emph{JHEP} {\bfseries 09}
  (2014) 051} [\href{https://arxiv.org/abs/1404.7517}{{\ttfamily 1404.7517}}].

\bibitem{gurau2016regular}
R.~G. Gurau and G.~Schaeffer, \emph{Regular colored graphs of positive degree},
  {\emph{Annales de l’Institut Henri Poincar{\'e} D} {\bfseries 3} (2016)
  257}.

\bibitem{Gurau:2015tua}
R.~Gurau, A.~Tanasa and D.~R. Youmans, \emph{{The double scaling limit of the
  multi-orientable tensor model}},
  \href{https://doi.org/10.1209/0295-5075/111/21002}{\emph{EPL} {\bfseries 111}
  (2015) 21002} [\href{https://arxiv.org/abs/1505.00586}{{\ttfamily
  1505.00586}}].

\bibitem{Bonzom:2019yik}
V.~Bonzom, V.~Nador and A.~Tanasa, \emph{{Diagrammatics of the quartic
  $O(N)^3$-invariant Sachdev-Ye-Kitaev-like tensor model}},
  \href{https://doi.org/10.1063/1.5095248}{\emph{J. Math. Phys.} {\bfseries 60}
  (2019) 072302} [\href{https://arxiv.org/abs/1903.01723}{{\ttfamily
  1903.01723}}].

\bibitem{Bonzom:2021kjy}
V.~Bonzom, V.~Nador and A.~Tanasa, \emph{{Double scaling limit for the
  O(N)$^{3}$-invariant tensor model}},
  \href{https://doi.org/10.1088/1751-8121/ac4898}{\emph{J. Phys. A} {\bfseries
  55} (2022) 135201} [\href{https://arxiv.org/abs/2109.07238}{{\ttfamily
  2109.07238}}].

\bibitem{Krajewski:2023tbv}
T.~Krajewski, T.~Muller and A.~Tanasa, \emph{{Double scaling limit of the
  prismatic tensor model}},
  \href{https://doi.org/10.1088/1751-8121/accf4e}{\emph{J. Phys. A} {\bfseries
  56} (2023) 235401} [\href{https://arxiv.org/abs/2301.02093}{{\ttfamily
  2301.02093}}].

\bibitem{Bonzom:2015axa}
V.~Bonzom, T.~Delepouve and V.~Rivasseau, \emph{{Enhancing non-melonic
  triangulations: A tensor model mixing melonic and planar maps}},
  \href{https://doi.org/10.1016/j.nuclphysb.2015.04.004}{\emph{Nucl. Phys. B}
  {\bfseries 895} (2015) 161}
  [\href{https://arxiv.org/abs/1502.01365}{{\ttfamily 1502.01365}}].

\bibitem{bonzom2017counting}
V.~Bonzom and L.~Lionni, \emph{Counting gluings of octahedra}, {\emph{The
  Electronic Journal of Combinatorics} {\bfseries 24} (2017) P3}.

\bibitem{Lionni:2017xvn}
L.~Lionni and J.~Th\"urigen, \emph{{Multi-critical behaviour of 4-dimensional
  tensor models up to order 6}},
  \href{https://doi.org/10.1016/j.nuclphysb.2019.02.026}{\emph{Nucl. Phys. B}
  {\bfseries 941} (2019) 600}
  [\href{https://arxiv.org/abs/1707.08931}{{\ttfamily 1707.08931}}].

\bibitem{Bonzom:2018btd}
V.~Bonzom, \emph{{Maximizing the number of edges in three-dimensional colored
  triangulations whose building blocks are balls}},
  \href{https://arxiv.org/abs/1802.06419}{{\ttfamily 1802.06419}}.

\bibitem{Ambjorn:2024pyv}
J.~Ambj\o{}rn and R.~Loll, \emph{{Causal Dynamical Triangulations: Gateway to
  Nonperturbative Quantum Gravity}},
  \href{https://arxiv.org/abs/2401.09399}{{\ttfamily 2401.09399}}.

\bibitem{Lionni:2019bzb}
L.~Lionni and J.-F. Marckert, \emph{{Iterated Foldings of Discrete Spaces and
  Their Limits: Candidates for the Role of Brownian Map in Higher Cimensions}},
  \href{https://doi.org/10.1007/s11040-021-09410-5}{\emph{Math. Phys. Anal.
  Geom.} {\bfseries 24} (2021) 39}
  [\href{https://arxiv.org/abs/1908.02259}{{\ttfamily 1908.02259}}].

\bibitem{Budd:2022dhq}
T.~Budd and L.~Lionni, \emph{{A family of triangulated 3-spheres constructed
  from trees}},  \href{https://arxiv.org/abs/2203.16105}{{\ttfamily
  2203.16105}}.

\bibitem{Delporte:2018iyf}
N.~Delporte and V.~Rivasseau, \emph{{The Tensor Track V: Holographic Tensors}},
   in \emph{{17th Hellenic School and Workshops on Elementary Particle Physics
  and Gravity}}, 4, 2018, \href{https://arxiv.org/abs/1804.11101}{{\ttfamily
  1804.11101}}.

\bibitem{Klebanov:2018fzb}
I.~R. Klebanov, F.~Popov and G.~Tarnopolsky, \emph{{TASI Lectures on Large $N$
  Tensor Models}}, \href{https://doi.org/10.22323/1.305.0004}{\emph{PoS}
  {\bfseries TASI2017} (2018) 004}
  [\href{https://arxiv.org/abs/1808.09434}{{\ttfamily 1808.09434}}].

\bibitem{Gurau:2019qag}
R.~G. Gurau, \emph{{Notes on tensor models and tensor field theories}},
  \href{https://doi.org/10.4171/aihpd/117}{\emph{Ann. Inst. H. Poincare D Comb.
  Phys. Interact.} {\bfseries 9} (2022) 159}
  [\href{https://arxiv.org/abs/1907.03531}{{\ttfamily 1907.03531}}].

\bibitem{Benedetti:2020seh}
D.~Benedetti, \emph{{Melonic CFTs}},
  \href{https://doi.org/10.22323/1.376.0168}{\emph{PoS} {\bfseries CORFU2019}
  (2020) 168} [\href{https://arxiv.org/abs/2004.08616}{{\ttfamily
  2004.08616}}].

\bibitem{Witten:2016iux}
E.~Witten, \emph{{An SYK-Like Model Without Disorder}},
  \href{https://doi.org/10.1088/1751-8121/ab3752}{\emph{J. Phys. A} {\bfseries
  52} (2019) 474002} [\href{https://arxiv.org/abs/1610.09758}{{\ttfamily
  1610.09758}}].

\bibitem{Klebanov:2016xxf}
I.~R. Klebanov and G.~Tarnopolsky, \emph{{Uncolored random tensors, melon
  diagrams, and the Sachdev-Ye-Kitaev models}},
  \href{https://doi.org/10.1103/PhysRevD.95.046004}{\emph{Phys. Rev. D}
  {\bfseries 95} (2017) 046004}
  [\href{https://arxiv.org/abs/1611.08915}{{\ttfamily 1611.08915}}].

\bibitem{Sachdev:1992fk}
S.~Sachdev and J.~Ye, \emph{{Gapless spin fluid ground state in a random,
  quantum Heisenberg magnet}},
  \href{https://doi.org/10.1103/PhysRevLett.70.3339}{\emph{Phys. Rev. Lett.}
  {\bfseries 70} (1993) 3339}
  [\href{https://arxiv.org/abs/cond-mat/9212030}{{\ttfamily
  cond-mat/9212030}}].

\bibitem{kitaev2015simple}
A.~Kitaev, \emph{A simple model of quantum holography},  Talks at KITP (2015).

\bibitem{Maldacena:2016hyu}
J.~Maldacena and D.~Stanford, \emph{{Remarks on the Sachdev-Ye-Kitaev model}},
  \href{https://doi.org/10.1103/PhysRevD.94.106002}{\emph{Phys. Rev. D}
  {\bfseries 94} (2016) 106002}
  [\href{https://arxiv.org/abs/1604.07818}{{\ttfamily 1604.07818}}].

\bibitem{Stanford:2017thb}
D.~Stanford and E.~Witten, \emph{{Fermionic Localization of the Schwarzian
  Theory}}, \href{https://doi.org/10.1007/JHEP10(2017)008}{\emph{JHEP}
  {\bfseries 10} (2017) 008}
  [\href{https://arxiv.org/abs/1703.04612}{{\ttfamily 1703.04612}}].

\bibitem{Maldacena:2016upp}
J.~Maldacena, D.~Stanford and Z.~Yang, \emph{{Conformal symmetry and its
  breaking in two dimensional Nearly Anti-de-Sitter space}},
  \href{https://doi.org/10.1093/ptep/ptw124}{\emph{PTEP} {\bfseries 2016}
  (2016) 12C104} [\href{https://arxiv.org/abs/1606.01857}{{\ttfamily
  1606.01857}}].

\bibitem{Mertens:2022irh}
T.~G. Mertens and G.~J. Turiaci, \emph{{Solvable models of quantum black holes:
  a review on Jackiw\textendash{}Teitelboim gravity}},
  \href{https://doi.org/10.1007/s41114-023-00046-1}{\emph{Living Rev. Rel.}
  {\bfseries 26} (2023) 4} [\href{https://arxiv.org/abs/2210.10846}{{\ttfamily
  2210.10846}}].

\bibitem{Bonzom:2017pqs}
V.~Bonzom, L.~Lionni and A.~Tanasa, \emph{{Diagrammatics of a colored SYK model
  and of an SYK-like tensor model, leading and next-to-leading orders}},
  \href{https://doi.org/10.1063/1.4983562}{\emph{J. Math. Phys.} {\bfseries 58}
  (2017) 052301} [\href{https://arxiv.org/abs/1702.06944}{{\ttfamily
  1702.06944}}].

\bibitem{Giombi:2017dtl}
S.~Giombi, I.~R. Klebanov and G.~Tarnopolsky, \emph{{Bosonic tensor models at
  large $N$ and small $\epsilon$}},
  \href{https://doi.org/10.1103/PhysRevD.96.106014}{\emph{Phys. Rev. D}
  {\bfseries 96} (2017) 106014}
  [\href{https://arxiv.org/abs/1707.03866}{{\ttfamily 1707.03866}}].

\bibitem{Benedetti:2021qyk}
D.~Benedetti, \emph{{Instability of complex CFTs with operators in the
  principal series}},
  \href{https://doi.org/10.1007/JHEP05(2021)004}{\emph{JHEP} {\bfseries 05}
  (2021) 004} [\href{https://arxiv.org/abs/2103.01813}{{\ttfamily
  2103.01813}}].

\bibitem{Benedetti:2019eyl}
D.~Benedetti, R.~Gurau and S.~Harribey, \emph{{Line of fixed points in a
  bosonic tensor model}},
  \href{https://doi.org/10.1007/JHEP06(2019)053}{\emph{JHEP} {\bfseries 06}
  (2019) 053} [\href{https://arxiv.org/abs/1903.03578}{{\ttfamily
  1903.03578}}].

\bibitem{Benedetti:2019ikb}
D.~Benedetti, R.~Gurau, S.~Harribey and K.~Suzuki, \emph{{Hints of unitarity at
  large $N$ in the $O(N)^3$ tensor field theory}},
  \href{https://doi.org/10.1007/JHEP02(2020)072}{\emph{JHEP} {\bfseries 02}
  (2020) 072} [\href{https://arxiv.org/abs/1909.07767}{{\ttfamily
  1909.07767}}].

\bibitem{Benedetti:2020yvb}
D.~Benedetti, R.~Gurau and K.~Suzuki, \emph{{Conformal symmetry and composite
  operators in the $O(N)^{3}$ tensor field theory}},
  \href{https://doi.org/10.1007/JHEP06(2020)113}{\emph{JHEP} {\bfseries 06}
  (2020) 113} [\href{https://arxiv.org/abs/2002.07652}{{\ttfamily
  2002.07652}}].

\bibitem{Harribey:2022esw}
S.~Harribey, \emph{{Renormalization in tensor field theory and the melonic
  fixed point}}, Ph.D. thesis, Heidelberg U., 2022.
\newblock \href{https://arxiv.org/abs/2207.05520}{{\ttfamily 2207.05520}}.
\newblock 10.11588/heidok.00031883.

\bibitem{Benedetti:2021wzt}
D.~Benedetti, R.~Gurau, S.~Harribey and D.~Lettera, \emph{{The F-theorem in the
  melonic limit}}, \href{https://doi.org/10.1007/JHEP02(2022)147}{\emph{JHEP}
  {\bfseries 02} (2022) 147}
  [\href{https://arxiv.org/abs/2111.11792}{{\ttfamily 2111.11792}}].

\bibitem{Berges:2023rqa}
J.~Berges, R.~Gurau and T.~Preis, \emph{{Asymptotic freedom in a strongly
  interacting scalar quantum field theory in four Euclidean dimensions}},
  \href{https://doi.org/10.1103/PhysRevD.108.016019}{\emph{Phys. Rev. D}
  {\bfseries 108} (2023) 016019}
  [\href{https://arxiv.org/abs/2301.09514}{{\ttfamily 2301.09514}}].

\bibitem{Peng:2016mxj}
C.~Peng, M.~Spradlin and A.~Volovich, \emph{{A Supersymmetric SYK-like Tensor
  Model}}, \href{https://doi.org/10.1007/JHEP05(2017)062}{\emph{JHEP}
  {\bfseries 05} (2017) 062}
  [\href{https://arxiv.org/abs/1612.03851}{{\ttfamily 1612.03851}}].

\bibitem{Choudhury:2017tax}
S.~Choudhury, A.~Dey, I.~Halder, L.~Janagal, S.~Minwalla and R.~Poojary,
  \emph{{Notes on melonic $O(N)^{q-1}$ tensor models}},
  \href{https://doi.org/10.1007/JHEP06(2018)094}{\emph{JHEP} {\bfseries 06}
  (2018) 094} [\href{https://arxiv.org/abs/1707.09352}{{\ttfamily
  1707.09352}}].

\bibitem{Bulycheva:2017ilt}
K.~Bulycheva, I.~R. Klebanov, A.~Milekhin and G.~Tarnopolsky, \emph{{Spectra of
  Operators in Large $N$ Tensor Models}},
  \href{https://doi.org/10.1103/PhysRevD.97.026016}{\emph{Phys. Rev. D}
  {\bfseries 97} (2018) 026016}
  [\href{https://arxiv.org/abs/1707.09347}{{\ttfamily 1707.09347}}].

\bibitem{Prakash:2017hwq}
S.~Prakash and R.~Sinha, \emph{{A Complex Fermionic Tensor Model in $d$
  Dimensions}}, \href{https://doi.org/10.1007/JHEP02(2018)086}{\emph{JHEP}
  {\bfseries 02} (2018) 086}
  [\href{https://arxiv.org/abs/1710.09357}{{\ttfamily 1710.09357}}].

\bibitem{Benedetti:2017fmp}
D.~Benedetti, S.~Carrozza, R.~Gurau and A.~Sfondrini, \emph{{Tensorial
  Gross-Neveu models}},
  \href{https://doi.org/10.1007/JHEP01(2018)003}{\emph{JHEP} {\bfseries 01}
  (2018) 003} [\href{https://arxiv.org/abs/1710.10253}{{\ttfamily
  1710.10253}}].

\bibitem{Benedetti:2018goh}
D.~Benedetti and R.~Gurau, \emph{{2PI effective action for the SYK model and
  tensor field theories}},
  \href{https://doi.org/10.1007/JHEP05(2018)156}{\emph{JHEP} {\bfseries 05}
  (2018) 156} [\href{https://arxiv.org/abs/1802.05500}{{\ttfamily
  1802.05500}}].

\bibitem{Benedetti:2018ghn}
D.~Benedetti and N.~Delporte, \emph{{Phase diagram and fixed points of
  tensorial Gross-Neveu models in three dimensions}},
  \href{https://doi.org/10.1007/JHEP01(2019)218}{\emph{JHEP} {\bfseries 01}
  (2019) 218} [\href{https://arxiv.org/abs/1810.04583}{{\ttfamily
  1810.04583}}].

\bibitem{Klebanov:2018nfp}
I.~R. Klebanov, A.~Milekhin, F.~Popov and G.~Tarnopolsky, \emph{{Spectra of
  eigenstates in fermionic tensor quantum mechanics}},
  \href{https://doi.org/10.1103/PhysRevD.97.106023}{\emph{Phys. Rev. D}
  {\bfseries 97} (2018) 106023}
  [\href{https://arxiv.org/abs/1802.10263}{{\ttfamily 1802.10263}}].

\bibitem{Giombi:2018qgp}
S.~Giombi, I.~R. Klebanov, F.~Popov, S.~Prakash and G.~Tarnopolsky,
  \emph{{Prismatic Large $N$ Models for Bosonic Tensors}},
  \href{https://doi.org/10.1103/PhysRevD.98.105005}{\emph{Phys. Rev. D}
  {\bfseries 98} (2018) 105005}
  [\href{https://arxiv.org/abs/1808.04344}{{\ttfamily 1808.04344}}].

\bibitem{Kim:2019upg}
J.~Kim, I.~R. Klebanov, G.~Tarnopolsky and W.~Zhao, \emph{{Symmetry Breaking in
  Coupled SYK or Tensor Models}},
  \href{https://doi.org/10.1103/PhysRevX.9.021043}{\emph{Phys. Rev. X}
  {\bfseries 9} (2019) 021043}
  [\href{https://arxiv.org/abs/1902.02287}{{\ttfamily 1902.02287}}].

\bibitem{Pakrouski:2018jcc}
K.~Pakrouski, I.~R. Klebanov, F.~Popov and G.~Tarnopolsky, \emph{{Spectrum of
  Majorana Quantum Mechanics with $O(4)^3$ Symmetry}},
  \href{https://doi.org/10.1103/PhysRevLett.122.011601}{\emph{Phys. Rev. Lett.}
  {\bfseries 122} (2019) 011601}
  [\href{https://arxiv.org/abs/1808.07455}{{\ttfamily 1808.07455}}].

\bibitem{Benedetti:2019rja}
D.~Benedetti, N.~Delporte, S.~Harribey and R.~Sinha, \emph{{Sextic tensor field
  theories in rank $3$ and $5$}},
  \href{https://doi.org/10.1007/JHEP06(2020)065}{\emph{JHEP} {\bfseries 06}
  (2020) 065} [\href{https://arxiv.org/abs/1912.06641}{{\ttfamily
  1912.06641}}].

\bibitem{Benedetti:2019sop}
D.~Benedetti and I.~Costa, \emph{{$SO(3)$-invariant phase of the $O(N)^3$
  tensor model}},
  \href{https://doi.org/10.1103/PhysRevD.101.086021}{\emph{Phys. Rev. D}
  {\bfseries 101} (2020) 086021}
  [\href{https://arxiv.org/abs/1912.07311}{{\ttfamily 1912.07311}}].

\bibitem{DeMelloKoch:2019tmo}
R.~De~Mello~Koch, D.~Gossman, N.~Hasina~Tahiridimbisoa and A.~L. Mahu,
  \emph{{Holography for Tensor models}},
  \href{https://doi.org/10.1103/PhysRevD.101.046004}{\emph{Phys. Rev. D}
  {\bfseries 101} (2020) 046004}
  [\href{https://arxiv.org/abs/1910.13982}{{\ttfamily 1910.13982}}].

\bibitem{Benedetti:2020iku}
D.~Benedetti and N.~Delporte, \emph{{Remarks on a melonic field theory with
  cubic interaction}},
  \href{https://doi.org/10.1007/JHEP04(2021)197}{\emph{JHEP} {\bfseries 04}
  (2021) 197} [\href{https://arxiv.org/abs/2012.12238}{{\ttfamily
  2012.12238}}].

\bibitem{Benedetti:2020sye}
D.~Benedetti, R.~Gurau and S.~Harribey, \emph{{Trifundamental quartic model}},
  \href{https://doi.org/10.1103/PhysRevD.103.046018}{\emph{Phys. Rev. D}
  {\bfseries 103} (2021) 046018}
  [\href{https://arxiv.org/abs/2011.11276}{{\ttfamily 2011.11276}}].

\bibitem{Harribey:2021xgh}
S.~Harribey, \emph{{Sextic tensor model in rank 3 at next-to-leading order}},
  \href{https://doi.org/10.1007/JHEP10(2022)037}{\emph{JHEP} {\bfseries 10}
  (2022) 037} [\href{https://arxiv.org/abs/2109.08034}{{\ttfamily
  2109.08034}}].

\bibitem{Jepsen:2023pzm}
C.~Jepsen and Y.~Oz, \emph{{RG flows and fixed points of O(N)$^{r}$ models}},
  \href{https://doi.org/10.1007/JHEP02(2024)035}{\emph{JHEP} {\bfseries 02}
  (2024) 035} [\href{https://arxiv.org/abs/2311.09039}{{\ttfamily
  2311.09039}}].

\bibitem{ponzano1968semiclassical}
G.~Ponzano and T.~E. Regge, \emph{Semiclassical limit of racah coefficients},
  in \emph{Spectroscopic and group theoretical methods in physics. Racah
  memorial volume}, pp.~1--58, North-Holland Publishing Co., (1968).

\bibitem{BenGeloun:2010qkf}
J.~Ben~Geloun, R.~Gurau and V.~Rivasseau, \emph{{EPRL/FK Group Field Theory}},
  \href{https://doi.org/10.1209/0295-5075/92/60008}{\emph{EPL} {\bfseries 92}
  (2010) 60008} [\href{https://arxiv.org/abs/1008.0354}{{\ttfamily
  1008.0354}}].

\bibitem{Baratin:2010wi}
A.~Baratin and D.~Oriti, \emph{{Group field theory with non-commutative metric
  variables}},
  \href{https://doi.org/10.1103/PhysRevLett.105.221302}{\emph{Phys. Rev. Lett.}
  {\bfseries 105} (2010) 221302}
  [\href{https://arxiv.org/abs/1002.4723}{{\ttfamily 1002.4723}}].

\bibitem{Baratin:2011tg}
A.~Baratin, F.~Girelli and D.~Oriti, \emph{{Diffeomorphisms in group field
  theories}}, \href{https://doi.org/10.1103/PhysRevD.83.104051}{\emph{Phys.
  Rev. D} {\bfseries 83} (2011) 104051}
  [\href{https://arxiv.org/abs/1101.0590}{{\ttfamily 1101.0590}}].

\bibitem{Oriti:2013aqa}
D.~Oriti, \emph{{Group field theory as the 2nd quantization of Loop Quantum
  Gravity}}, \href{https://doi.org/10.1088/0264-9381/33/8/085005}{\emph{Class.
  Quant. Grav.} {\bfseries 33} (2016) 085005}
  [\href{https://arxiv.org/abs/1310.7786}{{\ttfamily 1310.7786}}].

\bibitem{Jercher:2022mky}
A.~F. Jercher, D.~Oriti and A.~G.~A. Pithis, \emph{{Complete Barrett-Crane
  model and its causal structure}},
  \href{https://doi.org/10.1103/PhysRevD.106.066019}{\emph{Phys. Rev. D}
  {\bfseries 106} (2022) 066019}
  [\href{https://arxiv.org/abs/2206.15442}{{\ttfamily 2206.15442}}].

\bibitem{Asante:2022dnj}
S.~K. Asante, B.~Dittrich and S.~Steinhaus, \emph{{Spin Foams, Refinement
  Limit, and Renormalization}},
  \href{https://arxiv.org/abs/2211.09578}{{\ttfamily 2211.09578}}.

\bibitem{Perez:2012wv}
A.~Perez, \emph{{The Spin Foam Approach to Quantum Gravity}},
  \href{https://doi.org/10.12942/lrr-2013-3}{\emph{Living Rev. Rel.} {\bfseries
  16} (2013) 3} [\href{https://arxiv.org/abs/1205.2019}{{\ttfamily
  1205.2019}}].

\bibitem{Livine:2024hhc}
E.~R. Livine, \emph{{Spinfoam Models for Quantum Gravity: Overview}},
  \href{https://arxiv.org/abs/2403.09364}{{\ttfamily 2403.09364}}.

\bibitem{Freidel:2009hd}
L.~Freidel, R.~Gurau and D.~Oriti, \emph{{Group field theory renormalization -
  the 3d case: Power counting of divergences}},
  \href{https://doi.org/10.1103/PhysRevD.80.044007}{\emph{Phys. Rev. D}
  {\bfseries 80} (2009) 044007}
  [\href{https://arxiv.org/abs/0905.3772}{{\ttfamily 0905.3772}}].

\bibitem{Magnen:2009at}
J.~Magnen, K.~Noui, V.~Rivasseau and M.~Smerlak, \emph{{Scaling behaviour of
  three-dimensional group field theory}},
  \href{https://doi.org/10.1088/0264-9381/26/18/185012}{\emph{Class. Quant.
  Grav.} {\bfseries 26} (2009) 185012}
  [\href{https://arxiv.org/abs/0906.5477}{{\ttfamily 0906.5477}}].

\bibitem{Krajewski:2010yq}
T.~Krajewski, J.~Magnen, V.~Rivasseau, A.~Tanasa and P.~Vitale, \emph{{Quantum
  Corrections in the Group Field Theory Formulation of the EPRL/FK Models}},
  \href{https://doi.org/10.1103/PhysRevD.82.124069}{\emph{Phys. Rev. D}
  {\bfseries 82} (2010) 124069}
  [\href{https://arxiv.org/abs/1007.3150}{{\ttfamily 1007.3150}}].

\bibitem{Bonzom:2010ar}
V.~Bonzom and M.~Smerlak, \emph{{Bubble divergences from cellular cohomology}},
  \href{https://doi.org/10.1007/s11005-010-0414-4}{\emph{Lett. Math. Phys.}
  {\bfseries 93} (2010) 295} [\href{https://arxiv.org/abs/1004.5196}{{\ttfamily
  1004.5196}}].

\bibitem{Bonzom:2010zh}
V.~Bonzom and M.~Smerlak, \emph{{Bubble divergences from twisted cohomology}},
  \href{https://doi.org/10.1007/s00220-012-1477-0}{\emph{Commun. Math. Phys.}
  {\bfseries 312} (2012) 399}
  [\href{https://arxiv.org/abs/1008.1476}{{\ttfamily 1008.1476}}].

\bibitem{Bonzom:2011br}
V.~Bonzom and M.~Smerlak, \emph{{Bubble divergences: sorting out topology from
  cell structure}},
  \href{https://doi.org/10.1007/s00023-011-0127-y}{\emph{Annales Henri
  Poincare} {\bfseries 13} (2012) 185}
  [\href{https://arxiv.org/abs/1103.3961}{{\ttfamily 1103.3961}}].

\bibitem{Carrozza:2011jn}
S.~Carrozza and D.~Oriti, \emph{{Bounding bubbles: the vertex representation of
  3d Group Field Theory and the suppression of pseudo-manifolds}},
  \href{https://doi.org/10.1103/PhysRevD.85.044004}{\emph{Phys. Rev. D}
  {\bfseries 85} (2012) 044004}
  [\href{https://arxiv.org/abs/1104.5158}{{\ttfamily 1104.5158}}].

\bibitem{Carrozza:2012kt}
S.~Carrozza and D.~Oriti, \emph{{Bubbles and jackets: new scaling bounds in
  topological group field theories}},
  \href{https://doi.org/10.1007/JHEP06(2012)092}{\emph{JHEP} {\bfseries 06}
  (2012) 092} [\href{https://arxiv.org/abs/1203.5082}{{\ttfamily 1203.5082}}].

\bibitem{Baratin:2013rja}
A.~Baratin, S.~Carrozza, D.~Oriti, J.~Ryan and M.~Smerlak, \emph{{Melonic phase
  transition in group field theory}},
  \href{https://doi.org/10.1007/s11005-014-0699-9}{\emph{Lett. Math. Phys.}
  {\bfseries 104} (2014) 1003}
  [\href{https://arxiv.org/abs/1307.5026}{{\ttfamily 1307.5026}}].

\bibitem{BenGeloun:2011rc}
J.~Ben~Geloun and V.~Rivasseau, \emph{{A Renormalizable 4-Dimensional Tensor
  Field Theory}},
  \href{https://doi.org/10.1007/s00220-012-1549-1}{\emph{Commun. Math. Phys.}
  {\bfseries 318} (2013) 69} [\href{https://arxiv.org/abs/1111.4997}{{\ttfamily
  1111.4997}}].

\bibitem{Rivasseau:2011hm}
V.~Rivasseau, \emph{{Quantum Gravity and Renormalization: The Tensor Track}},
  \href{https://doi.org/10.1063/1.4715396}{\emph{AIP Conf. Proc.} {\bfseries
  1444} (2012) 18} [\href{https://arxiv.org/abs/1112.5104}{{\ttfamily
  1112.5104}}].

\bibitem{BenGeloun:2012pu}
J.~Ben~Geloun and D.~O. Samary, \emph{{3D Tensor Field Theory: Renormalization
  and One-loop $\beta$-functions}},
  \href{https://doi.org/10.1007/s00023-012-0225-5}{\emph{Annales Henri
  Poincare} {\bfseries 14} (2013) 1599}
  [\href{https://arxiv.org/abs/1201.0176}{{\ttfamily 1201.0176}}].

\bibitem{BenGeloun:2011jnm}
J.~Ben~Geloun and V.~Bonzom, \emph{{Radiative corrections in the
  Boulatov-Ooguri tensor model: The 2-point function}},
  \href{https://doi.org/10.1007/s10773-011-0782-2}{\emph{Int. J. Theor. Phys.}
  {\bfseries 50} (2011) 2819}
  [\href{https://arxiv.org/abs/1101.4294}{{\ttfamily 1101.4294}}].

\bibitem{rivasseau1991}
V.~Rivasseau, \emph{From Perturbative to Constructive Renormalization}.
  Princeton University Press, 1991.

\bibitem{BenGeloun:2012yk}
J.~Ben~Geloun, \emph{{Two and four-loop $\beta$-functions of rank 4
  renormalizable tensor field theories}},
  \href{https://doi.org/10.1088/0264-9381/29/23/235011}{\emph{Class. Quant.
  Grav.} {\bfseries 29} (2012) 235011}
  [\href{https://arxiv.org/abs/1205.5513}{{\ttfamily 1205.5513}}].

\bibitem{BenGeloun:2012phv}
J.~Ben~Geloun and E.~R. Livine, \emph{{Some classes of renormalizable tensor
  models}}, \href{https://doi.org/10.1063/1.4818797}{\emph{J. Math. Phys.}
  {\bfseries 54} (2013) 082303}
  [\href{https://arxiv.org/abs/1207.0416}{{\ttfamily 1207.0416}}].

\bibitem{BenGeloun:2013vwi}
J.~Ben~Geloun, \emph{{Renormalizable Models in Rank $d\geq 2$ Tensorial Group
  Field Theory}},
  \href{https://doi.org/10.1007/s00220-014-2142-6}{\emph{Commun. Math. Phys.}
  {\bfseries 332} (2014) 117}
  [\href{https://arxiv.org/abs/1306.1201}{{\ttfamily 1306.1201}}].

\bibitem{BenGeloun:2017xbd}
J.~Ben~Geloun and R.~Toriumi, \emph{{Renormalizable enhanced tensor field
  theory: The quartic melonic case}},
  \href{https://doi.org/10.1063/1.5022438}{\emph{J. Math. Phys.} {\bfseries 59}
  (2018) 112303} [\href{https://arxiv.org/abs/1709.05141}{{\ttfamily
  1709.05141}}].

\bibitem{Geloun:2023oyd}
J.~B. Geloun and R.~Toriumi, \emph{{One-loop beta-functions of quartic enhanced
  tensor field theories}},
  \href{https://doi.org/10.1088/1751-8121/acfdde}{\emph{J. Phys. A} {\bfseries
  57} (2024) 015401} [\href{https://arxiv.org/abs/2303.09829}{{\ttfamily
  2303.09829}}].

\bibitem{Carrozza:2012uv}
S.~Carrozza, D.~Oriti and V.~Rivasseau, \emph{{Renormalization of Tensorial
  Group Field Theories: Abelian U(1) Models in Four Dimensions}},
  \href{https://doi.org/10.1007/s00220-014-1954-8}{\emph{Commun. Math. Phys.}
  {\bfseries 327} (2014) 603}
  [\href{https://arxiv.org/abs/1207.6734}{{\ttfamily 1207.6734}}].

\bibitem{Samary:2012bw}
D.~O. Samary and F.~Vignes-Tourneret, \emph{{Just Renormalizable TGFT's on
  $U(1)^{d}$ with Gauge Invariance}},
  \href{https://doi.org/10.1007/s00220-014-1930-3}{\emph{Commun. Math. Phys.}
  {\bfseries 329} (2014) 545}
  [\href{https://arxiv.org/abs/1211.2618}{{\ttfamily 1211.2618}}].

\bibitem{Carrozza:2013wda}
S.~Carrozza, D.~Oriti and V.~Rivasseau, \emph{{Renormalization of a SU(2)
  Tensorial Group Field Theory in Three Dimensions}},
  \href{https://doi.org/10.1007/s00220-014-1928-x}{\emph{Commun. Math. Phys.}
  {\bfseries 330} (2014) 581}
  [\href{https://arxiv.org/abs/1303.6772}{{\ttfamily 1303.6772}}].

\bibitem{carrozza2014tensorial}
S.~Carrozza, \emph{Tensorial methods and renormalization in Group Field
  Theories}. Springer Science \& Business Media, 2014,
  [\href{https://arxiv.org/abs/1310.3736}{{\ttfamily 1310.3736}}].

\bibitem{Carrozza:2016vsq}
S.~Carrozza, \emph{{Flowing in Group Field Theory Space: a Review}},
  \href{https://doi.org/10.3842/SIGMA.2016.070}{\emph{SIGMA} {\bfseries 12}
  (2016) 070} [\href{https://arxiv.org/abs/1603.01902}{{\ttfamily
  1603.01902}}].

\bibitem{Carrozza:2014rba}
S.~Carrozza, \emph{{Discrete renormalization group for SU(2) tensorial group
  field theory}}, \href{https://doi.org/10.4171/aihpd/15}{\emph{Ann. Inst. H.
  Poincare D Comb. Phys. Interact.} {\bfseries 2} (2015) 49}
  [\href{https://arxiv.org/abs/1407.4615}{{\ttfamily 1407.4615}}].

\bibitem{Rivasseau:2015ova}
V.~Rivasseau, \emph{{Why are tensor field theories asymptotically free?}},
  \href{https://doi.org/10.1209/0295-5075/111/60011}{\emph{EPL} {\bfseries 111}
  (2015) 60011} [\href{https://arxiv.org/abs/1507.04190}{{\ttfamily
  1507.04190}}].

\bibitem{Gurau:2013oqa}
R.~Gurau and V.~Rivasseau, \emph{{The Multiscale Loop Vertex Expansion}},
  \href{https://doi.org/10.1007/s00023-014-0370-0}{\emph{Annales Henri
  Poincare} {\bfseries 16} (2015) 1869}
  [\href{https://arxiv.org/abs/1312.7226}{{\ttfamily 1312.7226}}].

\bibitem{Rivasseau:2017hpg}
V.~Rivasseau, \emph{{Loop Vertex Expansion for Higher Order Interactions}},
  \href{https://doi.org/10.1007/s11005-017-1037-9}{\emph{Lett. Math. Phys.}
  {\bfseries 108} (2018) 1147}
  [\href{https://arxiv.org/abs/1702.07602}{{\ttfamily 1702.07602}}].

\bibitem{Delepouve:2014hfa}
T.~Delepouve and V.~Rivasseau, \emph{{Constructive Tensor Field Theory: The
  $T^4_3$ Model}},
  \href{https://doi.org/10.1007/s00220-016-2680-1}{\emph{Commun. Math. Phys.}
  {\bfseries 345} (2016) 477}
  [\href{https://arxiv.org/abs/1412.5091}{{\ttfamily 1412.5091}}].

\bibitem{Rivasseau:2017xbk}
V.~Rivasseau and F.~Vignes-Tourneret, \emph{{Constructive Tensor Field Theory:
  The ${T_{4}^{4}}$ Model}},
  \href{https://doi.org/10.1007/s00220-019-03369-9}{\emph{Commun. Math. Phys.}
  {\bfseries 366} (2019) 567}
  [\href{https://arxiv.org/abs/1703.06510}{{\ttfamily 1703.06510}}].

\bibitem{Rivasseau:2021rlt}
V.~Rivasseau and F.~Vignes-Tourneret, \emph{{Can we make sense out of ''Tensor
  Field Theory''?}},
  \href{https://doi.org/10.21468/SciPostPhysCore.4.4.029}{\emph{SciPost Phys.
  Core} {\bfseries 4} (2021) 029}
  [\href{https://arxiv.org/abs/2101.04970}{{\ttfamily 2101.04970}}].

\bibitem{Chandra:2023kpp}
A.~Chandra and L.~Ferdinand, \emph{{A Stochastic Analysis Approach to Tensor
  Field Theories}},  \href{https://arxiv.org/abs/2306.05305}{{\ttfamily
  2306.05305}}.

\bibitem{Raasakka:2013kaa}
M.~Raasakka and A.~Tanasa, \emph{{Combinatorial Hopf algebra for the Ben
  Geloun-Rivasseau tensor field theory}}, {\emph{Sem. Lothar. Combin.}
  {\bfseries 70} (2014) B70d}
  [\href{https://arxiv.org/abs/1306.1022}{{\ttfamily 1306.1022}}].

\bibitem{Avohou:2015sia}
R.~C. Avohou, V.~Rivasseau and A.~Tanasa, \emph{{Renormalization and Hopf
  algebraic structure of the five-dimensional quartic tensor field theory}},
  \href{https://doi.org/10.1088/1751-8113/48/48/485204}{\emph{J. Phys. A}
  {\bfseries 48} (2015) 485204}
  [\href{https://arxiv.org/abs/1507.03548}{{\ttfamily 1507.03548}}].

\bibitem{Thurigen:2021zwf}
J.~Th\"urigen, \emph{{Renormalization in combinatorially non-local field
  theories: the Hopf algebra of 2-graphs}},
  \href{https://doi.org/10.1007/s11040-021-09390-6}{\emph{Math. Phys. Anal.
  Geom.} {\bfseries 24} (2021) 19}
  [\href{https://arxiv.org/abs/2102.12453}{{\ttfamily 2102.12453}}].

\bibitem{Thurigen:2021ntr}
J.~Th\"urigen, \emph{{Renormalization in Combinatorially Non-Local Field
  Theories: the BPHZ Momentum Scheme}},
  \href{https://doi.org/10.3842/SIGMA.2021.094}{\emph{SIGMA} {\bfseries 17}
  (2021) 094} [\href{https://arxiv.org/abs/2103.01136}{{\ttfamily
  2103.01136}}].

\bibitem{Benedetti:2014qsa}
D.~Benedetti, J.~Ben~Geloun and D.~Oriti, \emph{{Functional Renormalisation
  Group Approach for Tensorial Group Field Theory: a Rank-3 Model}},
  \href{https://doi.org/10.1007/JHEP03(2015)084}{\emph{JHEP} {\bfseries 03}
  (2015) 084} [\href{https://arxiv.org/abs/1411.3180}{{\ttfamily 1411.3180}}].

\bibitem{BenGeloun:2015xrk}
J.~Ben~Geloun, R.~Martini and D.~Oriti, \emph{{Functional Renormalization Group
  analysis of a Tensorial Group Field Theory on $\mathbb{R}^3$}},
  \href{https://doi.org/10.1209/0295-5075/112/31001}{\emph{EPL} {\bfseries 112}
  (2015) 31001} [\href{https://arxiv.org/abs/1508.01855}{{\ttfamily
  1508.01855}}].

\bibitem{Carrozza:2014rya}
S.~Carrozza, \emph{{Group field theory in dimension $4-\epsilon$}},
  \href{https://doi.org/10.1103/PhysRevD.91.065023}{\emph{Phys. Rev. D}
  {\bfseries 91} (2015) 065023}
  [\href{https://arxiv.org/abs/1411.5385}{{\ttfamily 1411.5385}}].

\bibitem{Benedetti:2015yaa}
D.~Benedetti and V.~Lahoche, \emph{{Functional Renormalization Group Approach
  for Tensorial Group Field Theory: A Rank-6 Model with Closure Constraint}},
  \href{https://doi.org/10.1088/0264-9381/33/9/095003}{\emph{Class. Quant.
  Grav.} {\bfseries 33} (2016) 095003}
  [\href{https://arxiv.org/abs/1508.06384}{{\ttfamily 1508.06384}}].

\bibitem{BenGeloun:2016rqa}
J.~Ben~Geloun, R.~Martini and D.~Oriti, \emph{{Functional Renormalisation Group
  analysis of Tensorial Group Field Theories on $\mathbb{R}^d$}},
  \href{https://doi.org/10.1103/PhysRevD.94.024017}{\emph{Phys. Rev. D}
  {\bfseries 94} (2016) 024017}
  [\href{https://arxiv.org/abs/1601.08211}{{\ttfamily 1601.08211}}].

\bibitem{Carrozza:2016tih}
S.~Carrozza and V.~Lahoche, \emph{{Asymptotic safety in three-dimensional SU(2)
  Group Field Theory: evidence in the local potential approximation}},
  \href{https://doi.org/10.1088/1361-6382/aa6d90}{\emph{Class. Quant. Grav.}
  {\bfseries 34} (2017) 115004}
  [\href{https://arxiv.org/abs/1612.02452}{{\ttfamily 1612.02452}}].

\bibitem{Carrozza:2017vkz}
S.~Carrozza, V.~Lahoche and D.~Oriti, \emph{{Renormalizable Group Field Theory
  beyond melonic diagrams: an example in rank four}},
  \href{https://doi.org/10.1103/PhysRevD.96.066007}{\emph{Phys. Rev. D}
  {\bfseries 96} (2017) 066007}
  [\href{https://arxiv.org/abs/1703.06729}{{\ttfamily 1703.06729}}].

\bibitem{Lahoche:2016xiq}
V.~Lahoche and D.~Ousmane~Samary, \emph{{Functional renormalization group for
  the U(1)-T$_5^6$ tensorial group field theory with closure constraint}},
  \href{https://doi.org/10.1103/PhysRevD.95.045013}{\emph{Phys. Rev. D}
  {\bfseries 95} (2017) 045013}
  [\href{https://arxiv.org/abs/1608.00379}{{\ttfamily 1608.00379}}].

\bibitem{BenGeloun:2018ekd}
J.~Ben~Geloun, T.~A. Koslowski, D.~Oriti and A.~D. Pereira, \emph{{Functional
  Renormalization Group analysis of rank 3 tensorial group field theory: The
  full quartic invariant truncation}},
  \href{https://doi.org/10.1103/PhysRevD.97.126018}{\emph{Phys. Rev. D}
  {\bfseries 97} (2018) 126018}
  [\href{https://arxiv.org/abs/1805.01619}{{\ttfamily 1805.01619}}].

\bibitem{Pithis:2020sxm}
A.~G.~A. Pithis and J.~Th\"urigen, \emph{{(No) phase transition in tensorial
  group field theory}},
  \href{https://doi.org/10.1016/j.physletb.2021.136215}{\emph{Phys. Lett. B}
  {\bfseries 816} (2021) 136215}
  [\href{https://arxiv.org/abs/2007.08982}{{\ttfamily 2007.08982}}].

\bibitem{Pithis:2020kio}
A.~G.~A. Pithis and J.~Th\"urigen, \emph{{Phase transitions in TGFT: functional
  renormalization group in the cyclic-melonic potential approximation and
  equivalence to O$(N)$ models}},
  \href{https://doi.org/10.1007/JHEP12(2020)159}{\emph{JHEP} {\bfseries 12}
  (2020) 159} [\href{https://arxiv.org/abs/2009.13588}{{\ttfamily
  2009.13588}}].

\bibitem{Geloun:2023ray}
J.~B. Geloun, A.~G.~A. Pithis and J.~Th\"urigen, \emph{{QFT with Tensorial and
  Local Degrees of Freedom: Phase Structure from Functional Renormalization}},
  \href{https://doi.org/10.1063/5.0158724}{\emph{J. Math. Phys.} {\bfseries 65}
  (2024) 032302} [\href{https://arxiv.org/abs/2305.06136}{{\ttfamily
  2305.06136}}].

\bibitem{Eichhorn:2017xhy}
A.~Eichhorn and T.~Koslowski, \emph{{Flowing to the continuum limit in tensor
  models for quantum gravity}},
  \href{https://doi.org/10.4171/aihpd/52}{\emph{Ann. Inst. H. Poincare D Comb.
  Phys. Interact.} {\bfseries 5} (2018) 173}
  [\href{https://arxiv.org/abs/1701.03029}{{\ttfamily 1701.03029}}].

\bibitem{Eichhorn:2018phj}
A.~Eichhorn, T.~Koslowski and A.~D. Pereira, \emph{{Status of
  background-independent coarse-graining in tensor models for quantum
  gravity}}, \href{https://doi.org/10.3390/universe5020053}{\emph{Universe}
  {\bfseries 5} (2019) 53} [\href{https://arxiv.org/abs/1811.12909}{{\ttfamily
  1811.12909}}].

\bibitem{Finocchiaro:2020fhl}
M.~Finocchiaro and D.~Oriti, \emph{{Renormalization of Group Field Theories for
  Quantum Gravity: New Computations and Some Suggestions}},
  \href{https://doi.org/10.3389/fphy.2020.552354}{\emph{Front. in Phys.}
  {\bfseries 8} (2021) 552354}
  [\href{https://arxiv.org/abs/2004.07361}{{\ttfamily 2004.07361}}].

\bibitem{Gielen:2013kla}
S.~Gielen, D.~Oriti and L.~Sindoni, \emph{{Cosmology from Group Field Theory
  Formalism for Quantum Gravity}},
  \href{https://doi.org/10.1103/PhysRevLett.111.031301}{\emph{Phys. Rev. Lett.}
  {\bfseries 111} (2013) 031301}
  [\href{https://arxiv.org/abs/1303.3576}{{\ttfamily 1303.3576}}].

\bibitem{Gielen:2013naa}
S.~Gielen, D.~Oriti and L.~Sindoni, \emph{{Homogeneous cosmologies as group
  field theory condensates}},
  \href{https://doi.org/10.1007/JHEP06(2014)013}{\emph{JHEP} {\bfseries 06}
  (2014) 013} [\href{https://arxiv.org/abs/1311.1238}{{\ttfamily 1311.1238}}].

\bibitem{Oriti:2016qtz}
D.~Oriti, L.~Sindoni and E.~Wilson-Ewing, \emph{{Emergent Friedmann dynamics
  with a quantum bounce from quantum gravity condensates}},
  \href{https://doi.org/10.1088/0264-9381/33/22/224001}{\emph{Class. Quant.
  Grav.} {\bfseries 33} (2016) 224001}
  [\href{https://arxiv.org/abs/1602.05881}{{\ttfamily 1602.05881}}].

\bibitem{Oriti:2015rwa}
D.~Oriti, D.~Pranzetti and L.~Sindoni, \emph{{Horizon entropy from quantum
  gravity condensates}},
  \href{https://doi.org/10.1103/PhysRevLett.116.211301}{\emph{Phys. Rev. Lett.}
  {\bfseries 116} (2016) 211301}
  [\href{https://arxiv.org/abs/1510.06991}{{\ttfamily 1510.06991}}].

\bibitem{Oriti:2018qty}
D.~Oriti, D.~Pranzetti and L.~Sindoni, \emph{{Black Holes as Quantum Gravity
  Condensates}}, \href{https://doi.org/10.1103/PhysRevD.97.066017}{\emph{Phys.
  Rev. D} {\bfseries 97} (2018) 066017}
  [\href{https://arxiv.org/abs/1801.01479}{{\ttfamily 1801.01479}}].

\bibitem{Gielen:2016dss}
S.~Gielen and L.~Sindoni, \emph{{Quantum Cosmology from Group Field Theory
  Condensates: a Review}},
  \href{https://doi.org/10.3842/SIGMA.2016.082}{\emph{SIGMA} {\bfseries 12}
  (2016) 082} [\href{https://arxiv.org/abs/1602.08104}{{\ttfamily
  1602.08104}}].

\bibitem{Oriti:2016acw}
D.~Oriti, \emph{{The universe as a quantum gravity condensate}},
  \href{https://doi.org/10.1016/j.crhy.2017.02.003}{\emph{Comptes Rendus
  Physique} {\bfseries 18} (2017) 235}
  [\href{https://arxiv.org/abs/1612.09521}{{\ttfamily 1612.09521}}].

\bibitem{Pithis:2019tvp}
A.~G.~A. Pithis and M.~Sakellariadou, \emph{{Group field theory condensate
  cosmology: An appetizer}},
  \href{https://doi.org/10.3390/universe5060147}{\emph{Universe} {\bfseries 5}
  (2019) 147} [\href{https://arxiv.org/abs/1904.00598}{{\ttfamily
  1904.00598}}].

\bibitem{Chirco:2017vhs}
G.~Chirco, D.~Oriti and M.~Zhang, \emph{{Group field theory and tensor
  networks: towards a Ryu\textendash{}Takayanagi formula in full quantum
  gravity}}, \href{https://doi.org/10.1088/1361-6382/aabf55}{\emph{Class.
  Quant. Grav.} {\bfseries 35} (2018) 115011}
  [\href{https://arxiv.org/abs/1701.01383}{{\ttfamily 1701.01383}}].

\bibitem{Chirco:2017wgl}
G.~Chirco, D.~Oriti and M.~Zhang, \emph{{Ryu-Takayanagi Formula for Symmetric
  Random Tensor Networks}},
  \href{https://doi.org/10.1103/PhysRevD.97.126002}{\emph{Phys. Rev. D}
  {\bfseries 97} (2018) 126002}
  [\href{https://arxiv.org/abs/1711.09941}{{\ttfamily 1711.09941}}].

\bibitem{Chirco:2019gsd}
G.~Chirco, \emph{{Holographic Entanglement in Group Field Theory}},
  \href{https://doi.org/10.3390/universe5100211}{\emph{Universe} {\bfseries 5}
  (2019) 211}.

\bibitem{Hayden:2016cfa}
P.~Hayden, S.~Nezami, X.-L. Qi, N.~Thomas, M.~Walter and Z.~Yang,
  \emph{{Holographic duality from random tensor networks}},
  \href{https://doi.org/10.1007/JHEP11(2016)009}{\emph{JHEP} {\bfseries 11}
  (2016) 009} [\href{https://arxiv.org/abs/1601.01694}{{\ttfamily
  1601.01694}}].

\bibitem{Colafranceschi:2020ern}
E.~Colafranceschi and D.~Oriti, \emph{{Quantum gravity states, entanglement
  graphs and second-quantized tensor networks}},
  \href{https://doi.org/10.1007/JHEP07(2021)052}{\emph{JHEP} {\bfseries 07}
  (2021) 052} [\href{https://arxiv.org/abs/2012.12622}{{\ttfamily
  2012.12622}}].

\bibitem{Jercher:2021bie}
A.~F. Jercher, D.~Oriti and A.~G.~A. Pithis, \emph{{Emergent cosmology from
  quantum gravity in the Lorentzian Barrett-Crane tensorial group field theory
  model}}, \href{https://doi.org/10.1088/1475-7516/2022/01/050}{\emph{JCAP}
  {\bfseries 01} (2022) 050}
  [\href{https://arxiv.org/abs/2112.00091}{{\ttfamily 2112.00091}}].

\bibitem{Marchetti:2022igl}
L.~Marchetti, D.~Oriti, A.~G.~A. Pithis and J.~Th\"urigen, \emph{{Phase
  transitions in TGFT: a Landau-Ginzburg analysis of Lorentzian quantum
  geometric models}},
  \href{https://doi.org/10.1007/JHEP02(2023)074}{\emph{JHEP} {\bfseries 02}
  (2023) 074} [\href{https://arxiv.org/abs/2209.04297}{{\ttfamily
  2209.04297}}].

\bibitem{Marchetti:2022nrf}
L.~Marchetti, D.~Oriti, A.~G.~A. Pithis and J.~Th\"urigen, \emph{{Mean-Field
  Phase Transitions in Tensorial Group Field Theory Quantum Gravity}},
  \href{https://doi.org/10.1103/PhysRevLett.130.141501}{\emph{Phys. Rev. Lett.}
  {\bfseries 130} (2023) 141501}
  [\href{https://arxiv.org/abs/2211.12768}{{\ttfamily 2211.12768}}].

\bibitem{Nador:2023inw}
V.~Nador, D.~Oriti, X.~Pang, A.~Tanasa and Y.-L. Wang, \emph{{Generalized
  Amit-Roginsky model from perturbations of 3D quantum gravity}},
  \href{https://doi.org/10.1103/PhysRevD.109.066008}{\emph{Phys. Rev. D}
  {\bfseries 109} (2024) 066008}
  [\href{https://arxiv.org/abs/2307.14211}{{\ttfamily 2307.14211}}].

\end{thebibliography}\endgroup


\end{document}